%% file: rev.tex
\RequirePackage{amsmath}
\documentclass[nofootinbib,floatfix,prd,aps,twocolumn,10pt]{revtex4-1}

\newcommand{\Tr}{\mathrm{Tr}}
\newcommand{\eps}{\varepsilon}

\usepackage{color}

\usepackage[utf8]{inputenc}
\usepackage{hyperref}
\let\oldvec\vec
\usepackage{amsmath}
\let\vec\oldvec
\usepackage{bbold}
\usepackage{amssymb}
\usepackage{graphicx}
\usepackage{mciteplus}
\usepackage[absolute]{textpos}

\begin{document}

\title{Dynamical stabilisation of complex Langevin simulations of QCD}
\author{Felipe Attanasio}
\email{pyfelipe@uw.edu}
\affiliation{Department of Physics, University of Washington, Box 351560,
Seattle, WA 98195, USA}
\author{Benjamin J\"ager}
\email{jaeger@cp3.sdu.dk}
\affiliation{CP3-Origins \& Danish IAS, Department of Mathematics and Computer
Science, University of Southern Denmark, 5230 Odense M, Denmark}
\begin{abstract}
	The ability to describe strongly interacting matter at finite temperature and
	baryon density provides the means to determine, for instance, the equation of
	state of QCD at non-zero baryon chemical potential. From a theoretical point of view,
	direct lattice simulations are hindered by the numerical sign problem, which prevents the use of traditional
	methods based on importance sampling. Despite recent successes, simulations
	using the complex Langevin method have been shown to exhibit instabilities, which
	cause convergence to wrong results. We introduce and discuss the method
	of Dynamic Stabilisation (DS), a modification of the complex Langevin process 
	aimed at solving these instabilities. We present results of DS being applied to
	the heavy-dense approximation of QCD, as well as QCD with staggered fermions at zero chemical potential and finite chemical potential at high temperature.
	Our findings show that DS can successfully deal with the aforementioned
	instabilities, opening the way for further progress. 
\begin{textblock}{20}(11.4,0.3)
	\begin{flushleft}
  CP3-Origins-2018-30 DNRF90\\
  NT@UW-18-05
  \end{flushleft}
 \end{textblock}% 
\end{abstract} %end of abstract
\maketitle

\input{1_intro}

\input{2_complexLangevin}

\input{5_dynamicStabilisation}

\input{6_alphaScan}

\input{7_dsBeta}

\input{8_deconfHD}

\input{9_staggered}

\input{10_conclusion}

\section{Acknowledgements}
	We would like to thank Gert Aarts, D\'enes Sexty, Erhard Seiler and Ion-Olimpiu Stamatescu for
	invaluable discussions and collaboration. We are indebted to Philippe de Forcrand for providing
	us with the HMC results for staggered fermions. We are grateful for the computing resources made
	available by HPC Wales. This work was facilitated though the use of advanced computational,
	storage, and networking infrastructure provided by the Hyak supercomputer system at the University of Washington.
	The work of FA was supported by US DOE Grant No. DE-FG02-97ER-41014.

\appendix

\input{3_hdqcd}

\input{4_staggeredIntro}
% BibTeX users please use
\bibliographystyle{style}
\bibliography{ref}
\end{document}

%% file: 1_intro.tex
\section{Introduction}
\label{intro}
Strongly interacting matter at finite baryon number density and temperature has been, and
remains, an active research subject to understand QCD under extreme conditions. Features of QCD are
typically studied in thermodynamic equilibrium, where the theory has two external parameters: the
temperature $T$ and and baryon chemical potential $\mu_B$. Varying those allows the exploration of the QCD
phase diagram in the $T$--$\mu_B$ plane. Known phases include ordinary nuclear matter and the
quark-gluon plasma (QGP), with a colour superconducting phase expected at large $\mu_B$. Of great
appeal are also the boundaries that mark the transition between these phases. This phase diagram has a fascinating structure, 
which is of significance for the study of hot and/or dense systems, such
as the early universe and heavy-ion collisions.

Heavy-ion collisions have been successfully used to investigate the high temperature behaviour of
QCD at the Relativistic Heavy Ion Collider (RHIC) and the Large Hadron Collider (LHC). These
facilities, together with future ones, namely the Facility for Antiproton and Ion Research (FAIR)
and the Nuclotron-based Ion Collider Facility (NICA), will further explore the phase diagram of QCD.
They will allow the study of hadronic interactions under extreme conditions, such as higher baryonic
density or very high temperatures.

From a theoretical perspective, some insight, at high temperature or density, can be gained from
perturbation theory. A full picture of the phase diagram, however, requires non-perturbative methods. Recent lattice results at non-zero temperature
include~\cite{Borsanyi:2013bia,Bazavov:2014pvz}. Typically, lattice QCD simulations at finite
baryon/quark density are carried out using the grand canonical ensemble, with the chemical potential
introduced as conjugate variable to the appropriate number density (quark, baryon, etc).
At finite quark chemical potential, the simulations have to overcome the infamous \textit{sign
problem}---a complex weight in the Euclidean path integral.
This imposes severe limitations on the applicability of standard numerical
methods~\cite{deForcrand:2010ys, Aarts:2015tyj}. Many approaches to deal with the sign problem have been proposed, including the
complex Langevin method~\cite{Parisi:1984cs,Klauder:1983nn,Klauder:1983zm,Klauder:1983sp},
strong coupling expansions~\cite{deForcrand:2014tha,Glesaaen:2015vtp,deForcrand:2017fky},
Lefschetz thimbles~\cite{Witten:2010cx,Cristoforetti:2012su,Alexandru:2015sua,Fujii:2013sra,Nishimura:2017vav},
holomorphic gradient flow~\cite{Alexandru:2016gsd}, 
density of states~\cite{Langfeld:2012ah,Gattringer:2016kco,Garron:2016noc,Endrodi:2018zda} 
and sign-optimized manifolds~\cite{Alexandru:2018fqp}.

The complex Langevin (CL) method is an extension of the stochastic quantisation
technique~\cite{Parisi:1980ys} to a complexified configuration space, without requiring a positive
weight~\cite{Parisi:1984cs,Klauder:1983zm,Klauder:1983sp}.
The complex nature of the method allows the circumvention of the sign problem, even when it is
severe~\cite{Aarts:2008wh,Aarts:2010gr,Aarts:2011zn}.
However, convergence to wrong limits has been observed both at Euclidean time~\cite{Ambjorn:1985iw, Ambjorn:1986fz, Aarts:2010aq}, and real time~\cite{Berges:2006xc, Berges:2007nr}.
These cases of incorrect convergence can be identified \textit{a posteriori}, based on the
theoretical justification of the method~\cite{Aarts:2009uq,Aarts:2011ax,Aarts:2012ft,Aarts:2013uza}.
Further discussions on the criteria for correct convergence of complex Langevin can be found in~\cite{Nagata:2016vkn,Nagata:2018net}.
Moreover, gauge cooling (GC)~\cite{Seiler:2012wz} has improved the convergence
of complex Langevin simulations for gauge theories.
The effects of gauge cooling on the complex Langevin method have been studied analytically in~\cite{Nagata:2015uga}.
Investigations of gauge cooling in random matrix theories has been performed in~\cite{Nagata:2016alq,Bloch:2017sex}.
 
Complex Langevin simulations, combined with gauge cooling, have successfully been used in QCD with a hopping expansion to all orders~\cite{Aarts:2014bwa}, with fully dynamical staggered fermions~\cite{Sexty:2013ica} and to map the phase diagram of QCD in the heavy-dense limit
(HDQCD)~\cite{Aarts:2016qrv}. In that work, we noticed that, despite the use of gauge cooling, instabilities might
appear during the simulations.
Here, we introduce and elaborate on our method of Dynamic Stabilisation (DS), which has
been constructed to deal with these instabilities. 

This paper is organised as follows: in section~\ref{sec.complexLangevin} we review the complex
Langevin method.
Section~\ref{sec.DS} motivates and introduces the method of dynamic stabilisation. 
Tests of this procedure, applied to QCD in the limit of heavy-dense quarks (HDQCD)~\cite{Bender:1992gn,Aarts:2008rr} are discussed in sections~\ref{sec.alphaScan} and~\ref{sec.betaScan}.
Section~\ref{sec.stag.DS} shows 
the outcome of applying dynamic stabilisation to simulations with staggered quarks at zero chemical
potential
and at finite chemical potential and high temperatures.
We summarise our findings in sec.~\ref{sec.summary}.
Appendices~\ref{sec.hdqcd} and~\ref{sec.stag} review the HDQCD approximation
and the staggered formulation of lattice quarks, which have been used in our investigations.

Preliminary results on dynamic stabilisation have already appeared in~\cite{Aarts:2016qhx,
Attanasio:2016mhc, Attanasio:2017rxk}.

%% file: 2_complexLangevin.tex
\section{Complex Langevin}\label{sec.complexLangevin}
We study QCD by employing the method of stochastic quantisation~\cite{Parisi:1980ys}, with which
quantum expectation values can be computed using Langevin dynamics. These expectation values are
evaluated as averages over a stochastic process, in which dynamical variables are evolved over a
fictitious time $\theta$. Notably, importance sampling does not enter in this formulation.
The partition function for lattice QCD in the grand canonical ensemble, where the (quark) chemical
potential $\mu$ couples to the quark number, is
\begin{equation}
	Z = \int DU \, \det M \, e^{-S_{\mathrm{YM}}}
	\equiv \int DU \, e^{-S}\,,
\end{equation}
where, for the second equality, the bilinear quark fields have been integrated out. $U$ represents the gauge links,
$S_{\mathrm{YM}}$ is the Yang--Mills action and $M$ the fermion matrix, which depends on the gauge
links and the chemical potential, and $S = S_{\mathrm{YM}} - \ln \det M$.

We consider a SU($3$) gauge theory with links $U_{x,\nu}$, defined on a lattice of spatial volume
$N_s^3$ and temporal extent $N_\tau$. A Langevin update, using a first-order discretisation scheme
in the Langevin time $\theta = n \eps$, is given by~\cite{Damgaard:1987rr}
\begin{equation}
	U_{x,\nu}(\theta + \eps) = \exp\left[ i\lambda^a \left(\eps K^a_{x,\nu} + \sqrt{\eps} \eta^a_{x,\nu} \right) \right] U_{x,\nu}(\theta)\,,
\end{equation}
where $\lambda^a$ are the Gell-Mann matrices, with $\Tr \left[ \lambda^a \lambda^b \right] = 2 \delta^{ab}$, and $\eta^a_{x,\nu}$ are Gaussian white noise fields satisfying
\begin{equation}
	\langle \eta^a_{x,\mu} \rangle = 0\,, \quad \langle \eta^a_{x,\mu} \eta^b_{y,\nu} \rangle = 2 \delta_{xy} \delta^{ab} \delta_{\mu\nu}\,.
\end{equation}
The Langevin drift, $K^a_{x,\nu}$, is obtained from the action $S$,
\begin{equation}\label{eq.drift}
	K^a_{x,\nu} = -D^a_{x,\nu} S = -D^a_{x,\nu} S_{\mathrm{YM}} + \Tr \left[ M^{-1} D^a_{x,\nu} M \right]\,,
\end{equation}
where $D^a_{x,\nu}$ is the gauge group derivative
\begin{equation}
	D^a_{x,\nu} f(U) = \frac{\partial}{\partial \alpha} f \left.\left( e^{i \alpha \lambda^a} U_{x,\nu} \right)\right|_{\alpha=0}\,.
\end{equation}

Poles may appear in the drift in the presence of quarks, when $\det M = 0$ and $M^{-1}$ does not exist.
In some situations this has a negative impact on the
results~\cite{Mollgaard:2013qra,Nishimura:2015pba}, but, as far as understood, this is not the case
in HDQCD~\cite{Aarts:2014bwa,Splittorff:2014zca}. For further reference, we refer the reader to the
extended discussion on the issues arising from the branch cuts of the 
logarithm of the determinant~\cite{Mollgaard:2013qra,Splittorff:2014zca,Greensite:2014cxa,Nishimura:2015pba}
In~\cite{Aarts:2017vrv} it was clarified that it is the drift's behaviour around the poles, rather than the branch cuts, that affects the reliability of the complex Langevin method.
It is necessary to employ adaptive algorithms to change the Langevin step size $\eps$, in order to
avoid numerical instabilities and regulate large values of the drift~\cite{Aarts:2009dg}.

When the sign problem is present, the Langevin drift is complex.
This results in the exploration of a larger configuration space.
The sign problem is circumvented by allowing the gauge links to take values in enlarged
manifolds~\cite{Aarts:2008rr,Parisi:1984cs,Klauder:1983nn,Klauder:1983zm,Klauder:1983sp,Aarts:2008wh,Aarts:2009uq}.
In the case of QCD, the gauge group extends from SU($3$) to SL($3,\mathbb{C}$).
The extra freedom can lead to trajectories where the non-unitary parts of the gauge
links are not small deformations of the original theory.
The ``distance'' from the unitary manifold can be used to identify these trajectories.
A possible measurement of this distance is given by the unitarity norm
\begin{equation}
	d = \frac{1}{3 \Omega} \sum_{x,\nu} \Tr \left[ U_{x,\nu} U^\dagger_{x,\nu} - \mathbb{1} \right]^2 \geq 0\,,
\end{equation}
where $\Omega = N_s^3 N_\tau$ is the four dimensional lattice volume.
This norm is invariant under SU($3$) gauge transformations and vanishes only if all links $U_{x,\nu}$ are unitary.

It has been shown that simulations in which the unitarity norm is kept under control lead to reliable results, matching exact ones or results from different methods, when available \cite{Seiler:2012wz,Aarts:2016qrv}.
One procedure to reduce the distance to the unitary manifold is known as gauge cooling \cite{Seiler:2012wz}.
It consists of a sequence of SL($3,\mathbb{C}$) gauge transformations, designed to decrease the unitarity norm in a steepest descent style
\begin{align}
	U_{x,\nu} &\to e^{-\eps \alpha \lambda^a f^a_x} \, U_{x,\nu} \, e^{\eps \alpha \lambda^a f^a_x}\,,\\
	f^a_x = 2 \sum_\nu &\Tr\left[ \lambda^a \left( U_{x,\nu} U^\dagger_{x,\nu} - U^\dagger_{x-\nu,\nu} U_{x-\nu,\nu} \right) \right]\,.
\end{align}
The transformation parameters, $f^a_x$, are obtained by requiring that the first variation of $d$ with respect to a gauge transformation is negative semi-definite.
The coefficient $\alpha$ can be changed adaptively to optimise the cooling procedure \cite{Aarts:2013uxa}.
A variable number of gauge cooling steps, depending on the rate of change of the unitarity norm, can be applied between successive Langevin steps \cite{Aarts:2015hnb}.

In our studies involving the heavy-dense limit of QCD (HDQCD)~\cite{Bender:1992gn,Aarts:2008rr}, we have
considered the expectation value of the traced (inverse) Polyakov loops,
\begin{align}
	\langle P \rangle &= \frac{1}{V} \sum_{\vec{x}} \langle P_{\vec{x}} \rangle\,, \qquad P_{\vec{x}} = \frac{1}{3} \Tr \mathcal{P}_{\vec{x}}\,,\\
	\langle P^{-1} \rangle &= \frac{1}{V} \sum_{\vec{x}} \langle P^{-1}_{\vec{x}} \rangle\,, \qquad P^{-1}_{\vec{x}} = \frac{1}{3} \Tr \mathcal{P}^{-1}_{\vec{x}}\,,
\end{align}
where $V$ is the spatial volume.
The average Polyakov loop is an order parameter for Yang-Mills theories, as it is related to
the free energy of a single quark by $\langle P \rangle \sim e^{-F_q / T}$.
In the presence of dynamical quarks, it is no longer an order parameter.
However, it still provides information on whether quarks are free or confined within hadrons.
Another useful observable is the average phase of the quark determinant, measured in a phase quenched ensemble,
\begin{equation}
	\left\langle e^{2i \phi} \right\rangle = \left\langle \frac{\det M(\mu)}{\det M(-\mu)} \right\rangle_{\mathrm{PQ}}\,.
\end{equation}
When the sign problem is mild, the phase is not expected to vary much, leading to an average close to unity.
On the other hand, in situations with severe sign problems, $e^{2i \phi}$ can average out to zero.
When dealing with fully dynamical quarks, we have studied the chiral condensate
\begin{equation}
	\langle \overline{\psi} \psi \rangle = \frac{T}{V} \frac{\partial}{\partial m} \ln Z\,.
\end{equation}
This is an order parameter only for massless quarks, but like the Polyakov loop, still provides information on quark confinement in general.

%% file: 5_dynamicStabilisation.tex
\section{Dynamic stabilisation}\label{sec.DS}
We found that even with a large number of gauge cooling steps, instabilities still may
appear~\cite{Aarts:2016qrv} in HDQCD simulations. These change the distribution of the observables during the Langevin
process and lead to wrong results.

\begin{figure}
	\centering
	\includegraphics[width=0.95\linewidth]{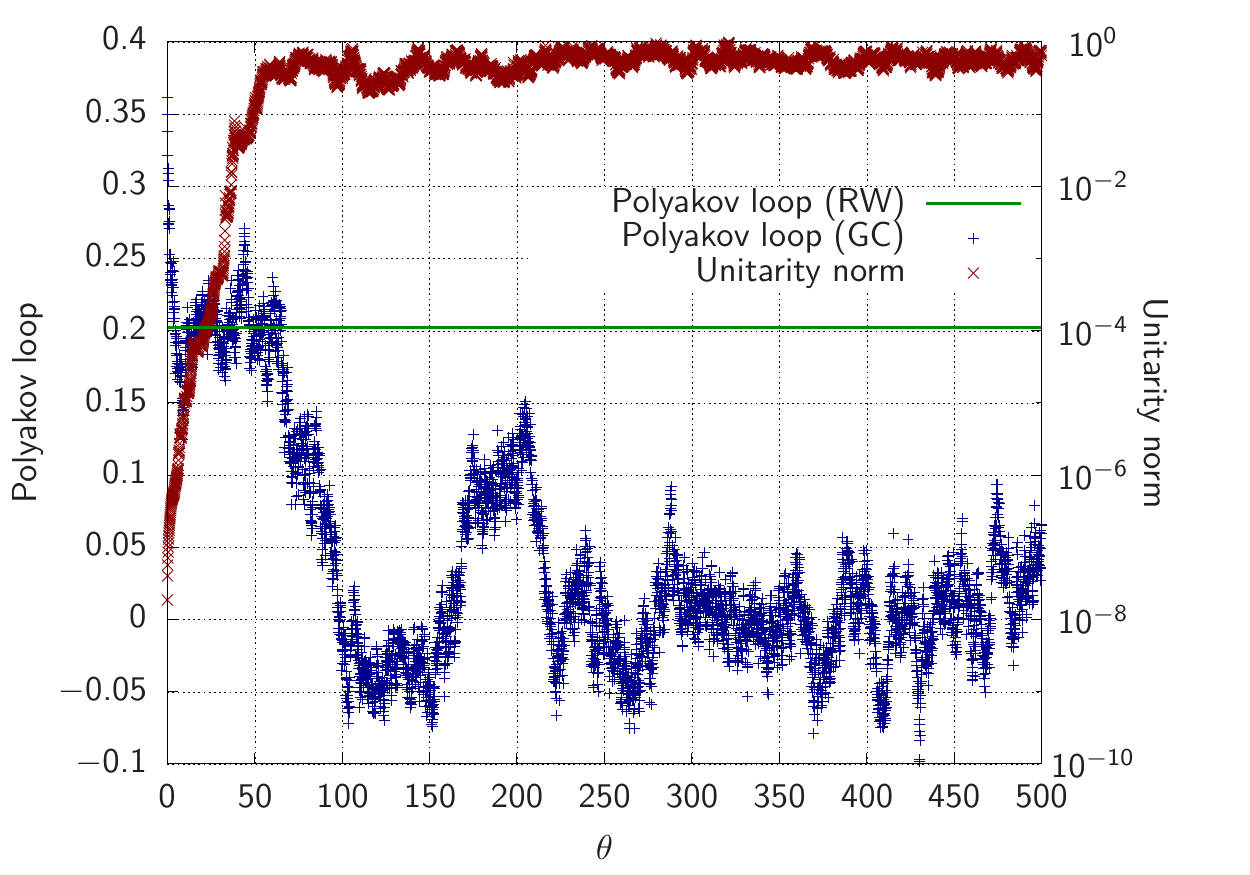} 
	\caption{\label{fig.HD.tunnel} The average Polyakov loop (red) and unitarity norm (green) as
	functions of the Langevin time for HDQCD on a $10^3 \times 4$ lattice, with $\kappa=0.04$,
	$\beta=5.8$ and $\mu=0.7$. Agreement with reweighting is found when the unitarity norm is
	lower than $O(0.1)$. The data was first presented in~\protect\cite{Aarts:2016qrv}.}
\end{figure}
Figure~\ref{fig.HD.tunnel} shows the Langevin time evolution of the Polyakov loop and of the
unitarity norm.
This situation has a very mild sign problem, with average phase $\langle e^{2i\phi} \rangle =
0.9978(2) - 0.0003(57)i$, and thus results from reweighting are reliable.
We observe two distinct regions: one is characterised by a sufficiently small
unitarity norm and agreement between gauge cooling and reweighting results. At a larger
Langevin time, i.e. $\theta \gtrsim 50$, the agreement disappears as the unitarity norm becomes too 
large. It has been concluded in~\cite{Aarts:2016qrv} that a large unitarity norm is an indicator of
these instabilities, with $0.03$ being a conservative threshold, after which results become unreliable.

To keep the unitarity norm under control we have developed a new technique---dynamic
stabilisation---which consists of adding a SU($3$) gauge invariant force to the Langevin drift.
This force is designed to grow rapidly with the unitarity
norm $d$ and to be directed towards the SU($3$) manifold. One possible implementation is given by the
substitution
\begin{equation}\label{eq.drift.lang.DS}
	K^a_{x,\mu} \to K^a_{x,\mu} + i \alpha_{\mathrm{DS}} M^a_x\,,
\end{equation}
with the new term
\begin{equation}\label{eq.DS.term}
	M^a_x = i b^a_x \left( \sum_c b^c_x b^c_x \right)^3
\end{equation}
and
\begin{equation}
	b^a_x = \Tr \left[ \lambda^a \sum_\nu U_{x,\nu} U^\dagger_{x,\nu} \right]\,.
\end{equation}
We remark that our choice for the additional force, which acts equally in all four directions, is
not unique. The parameter $\alpha_{\mathrm{DS}}$ allows us to control the strength of the
this force. A similar strategy has been used successfully for
nonrelativistic fermions in one dimension~\cite{Loheac:2017yar,Rammelmuller:2017vqn}.

We point out that $M^a_x$ is not invariant under general SL($3, \mathbb{C}$) gauge transformations,
but it is with respect to SU($3$) transformations. Moreover, it is not holomorphic, since it is constructed to be a function of only the 
non-unitary part of the gauge links, i.e., of the combination $UU^\dagger$.
This is necessary to make $M^a_x$ scale with the unitarity norm, such that explorations of the non-unitary directions can be controlled.
Therefore, it cannot be obtained from a derivative of the action.
This invalidates the standard justification for the validity of complex
Langevin~\cite{Aarts:2009uq,Aarts:2011ax}, which require a holomorphic Langevin drift.
Nevertheless, numerical evidence of the convergence to the correct limit of CL simulations with
dynamic stabilisation will be shown in sections~\ref{sec.betaScan} and~\ref{sec.stag.DS}.

A na\"ive expansion of $M^a_x$ in powers of the lattice spacing is possible if one writes the
SL($3,\mathbb{C}$) gauge links as $U_{x,\nu} = \exp\left[ i a \lambda^a \left( A^a_{x,\nu} + i
B^a_{x,\nu} \right) \right]$.
Then, formally,
\begin{align}
	M^a_x &\sim a^7 \left( \sum_c \overline{B}^c_x \overline{B}^c_x \right)^3 \overline{B}^a_x + O(a^8)\,,\\
	\overline{B}^a_x &= \sum_\nu B^a_{x,\nu}\,.
\end{align}
The continuum behaviour will be discussed in section \ref{sec.betaScan}, where we
show results for different gauge couplings. By construction, the DS
drift is purely imaginary and thus acts only on the imaginary parts of the Langevin drift.
Checks of how the Langevin and DS drifts behave in a situation with a severe sign problem are shown in section~\ref{sec.alphaScan}.

\begin{figure}
	\centering
	\includegraphics[width=0.85\linewidth]{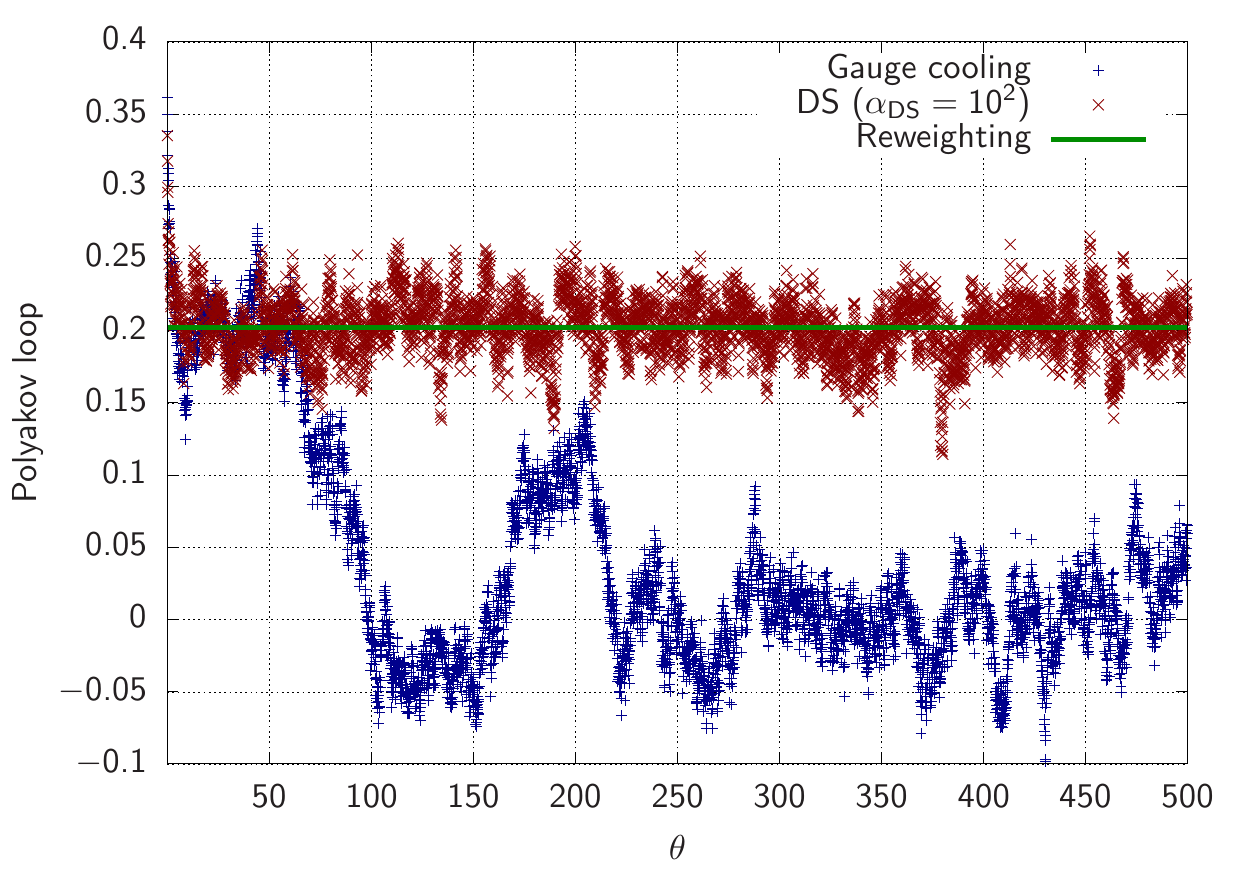}
	\caption{\label{fig.HD.tunnel.DS} The average Polyakov loop as function of the Langevin time for
	HDQCD on a $10^3 \times 4$ lattice,  with $\kappa=0.04$, $\beta=5.8$, $\mu=0.7$ and
	$\alpha_{\mathrm{DS}}=100$.}
\end{figure}

An initial result of complex Langevin simulations using one step of gauge cooling\footnote{We
have checked that multiple gauge cooling steps lead to a negligible improvement. 
At least one gauge cooling step is required, since dynamic stabilisation does not affect the real
part of the drift, which can develop large fluctuations~\cite{Aarts:2013uxa}.}
and dynamic stabilisation is shown in fig.~\ref{fig.HD.tunnel.DS}.
We have used the same parameters of fig.~\ref{fig.HD.tunnel} and found agreement with reweighting for
the entire length of the simulation.
Figure~\ref{fig.HD.tunnel.DS} also demonstrates that it is possible to stabilise complex Langevin
simulations in a way that gauge cooling alone is not able to, allowing for longer simulation times and thus smaller statistical errors.

%% file: 6_alphaScan.tex
\section{Dependence of observables on \texorpdfstring{$\alpha_{\mathrm{DS}}$}~}\label{sec.alphaScan}
The complexity of gauge theories makes it difficult to predict the effect of the
control parameter $\alpha_{\mathrm{DS}}$ on the Langevin dynamics.
However, two limiting cases can be expected: for small $\alpha_{\mathrm{DS}}$
the DS drift becomes very small, essentially not affecting the dynamics. For
large values of $\alpha_{\mathrm{DS}}$, the DS force heavily suppresses
excursions into the non-unitary directions of SL($3,\mathbb{C}$), which can be
interpreted as a gradual reunitarisation of the gauge links.
We illustrate the effect of different $\alpha_{\mathrm{DS}}$ on complex
Langevin simulations of HDQCD in two cases. The first scenario corresponds to an
average phase of the quark determinant close to unity, i.e., when the sign
problem is mild and comparisons with reweighting are possible. In the second
case, the average phase is very small, indicating a severe sign problem.
Both scenarios have been simulated with inverse coupling $\beta=5.8$ and hopping
parameter $\kappa=0.04$. Additionally, one gauge cooling step has been
applied between consecutive Langevin updates.

\begin{figure}
	\centering
	\includegraphics[width=0.85\linewidth]{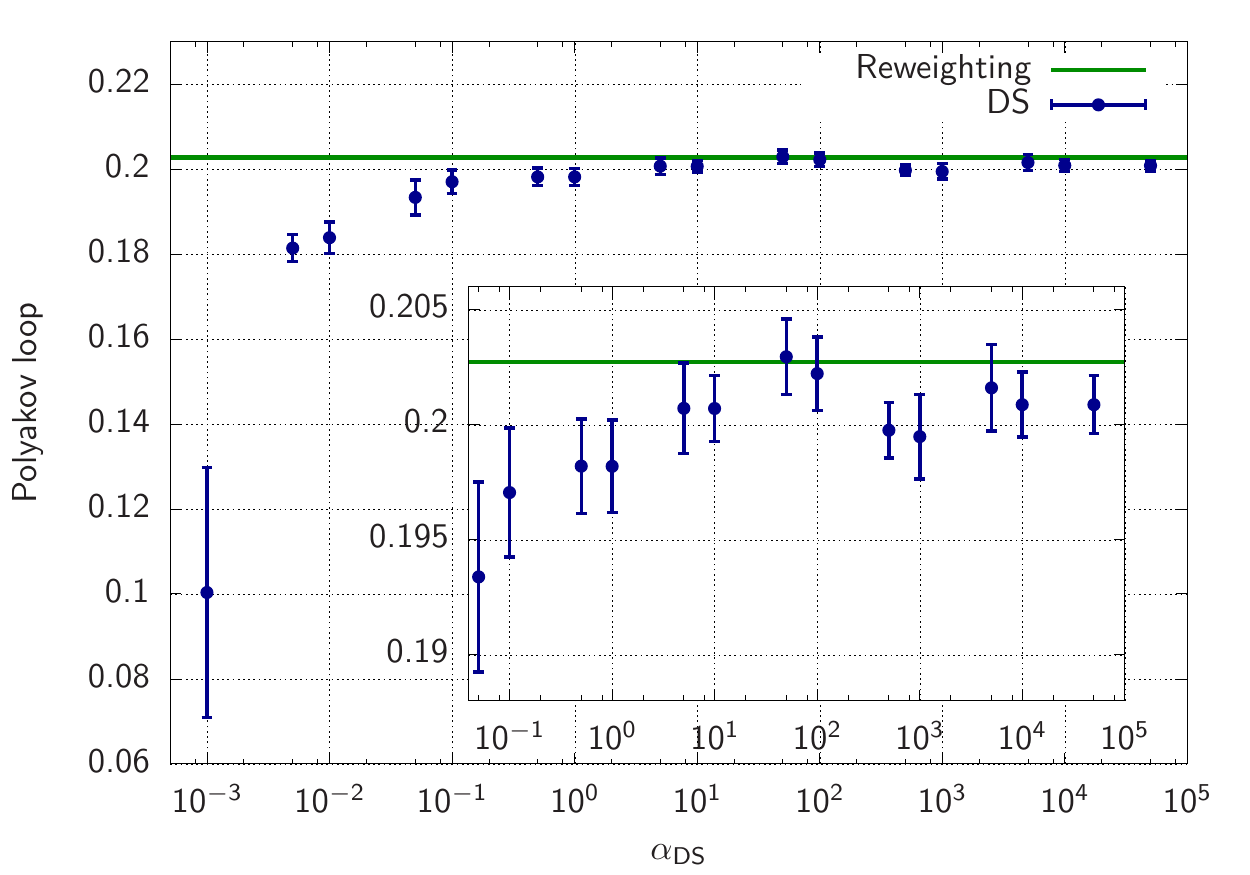}
	\caption{\label{fig.alpha.scan} The average Polyakov loop as a function of
	$\alpha_{\mathrm{DS}}$ compared with the result generated with reweighting for
	HDQCD in a $10^3 \times 4$ lattice, with $\kappa=0.04$, $\beta=5.8$ and
	$\mu=0.7$. Agreement is found once $\alpha_{\mathrm{DS}}$ is sufficiently
	large.}
\end{figure}

Results for the first scenario are shown in fig.~\ref{fig.alpha.scan}. These simulations use a
volume of $\Omega=10^3 \times 4$ and chemical potential $\mu=0.7$.
We find agreement with reweighting for sufficiently large
$\alpha_{\mathrm{DS}}$.
Figure~\ref{fig.alpha.scan} seems to indicate that one could choose an
arbitrarily large $\alpha_{\mathrm{DS}}$. However, it is
necessary to keep in mind that, for these parameters, these simulations have the
average phase of the fermion determinant close to unity, and a very mild sign
problem.

\begin{figure}
	\centering
	\includegraphics[width=0.85\linewidth]{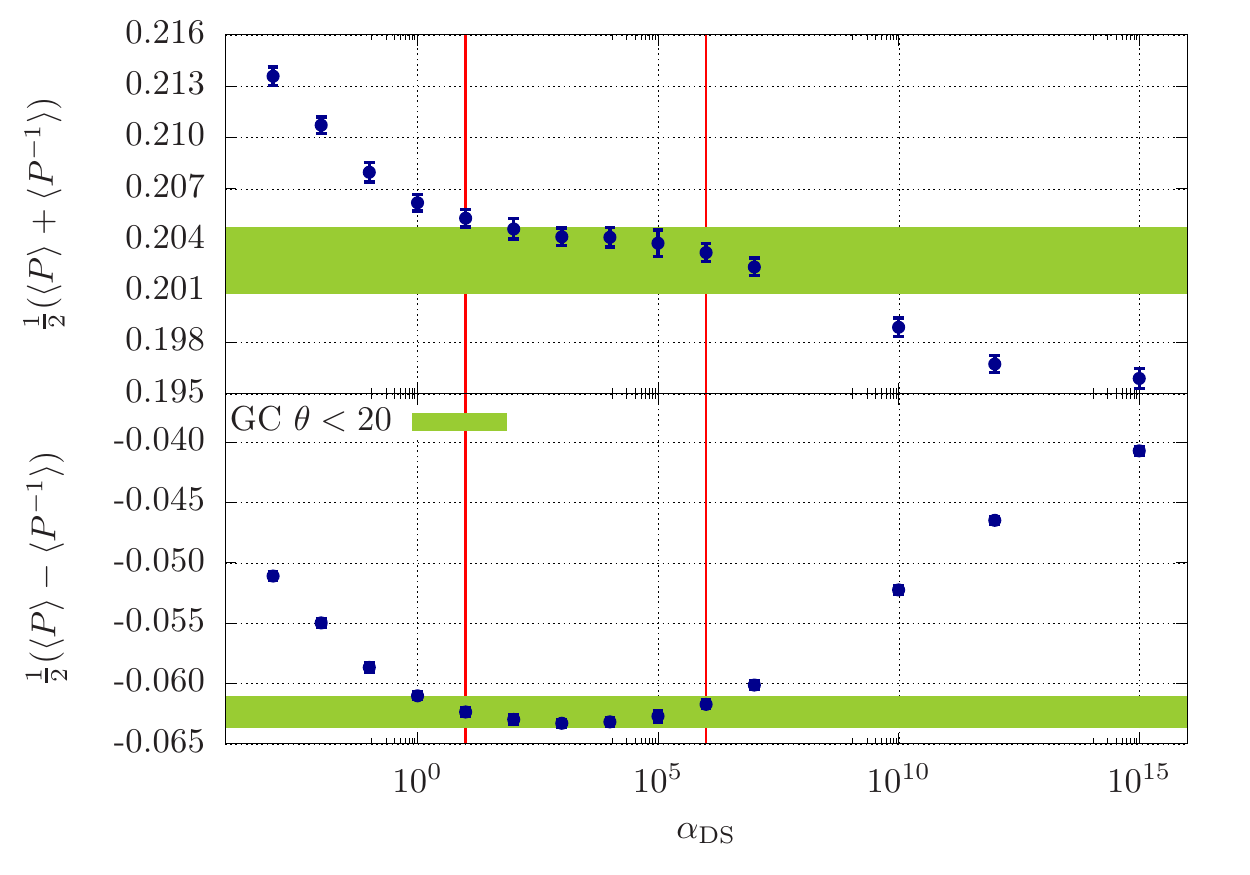}
	\caption{\label{fig.alpha.scan.det.zero}The average Polyakov loop as function
	of $\alpha_{\mathrm{DS}}$ at $N_\tau=20$ and $\mu=2.45$. Also shown are results
	from simulations with gauge cooling. The vertical red
	lines indicate the region where there is agreement between GC and DS runs.}
\end{figure}

When the sign problem is severe, indicated by a highly oscillating phase of the
fermion determinant, reweighting cannot be applied reliably. 
In principle, complex Langevin combined with gauge cooling is a viable option to simulate these
regions of the phase diagram. However, as seen in fig.~\ref{fig.HD.tunnel}, we 
have observed disagreement with reweighting, when the unitarity norm becomes too large. 
\begin{figure}
	\centering
	\includegraphics[width=0.85\linewidth]{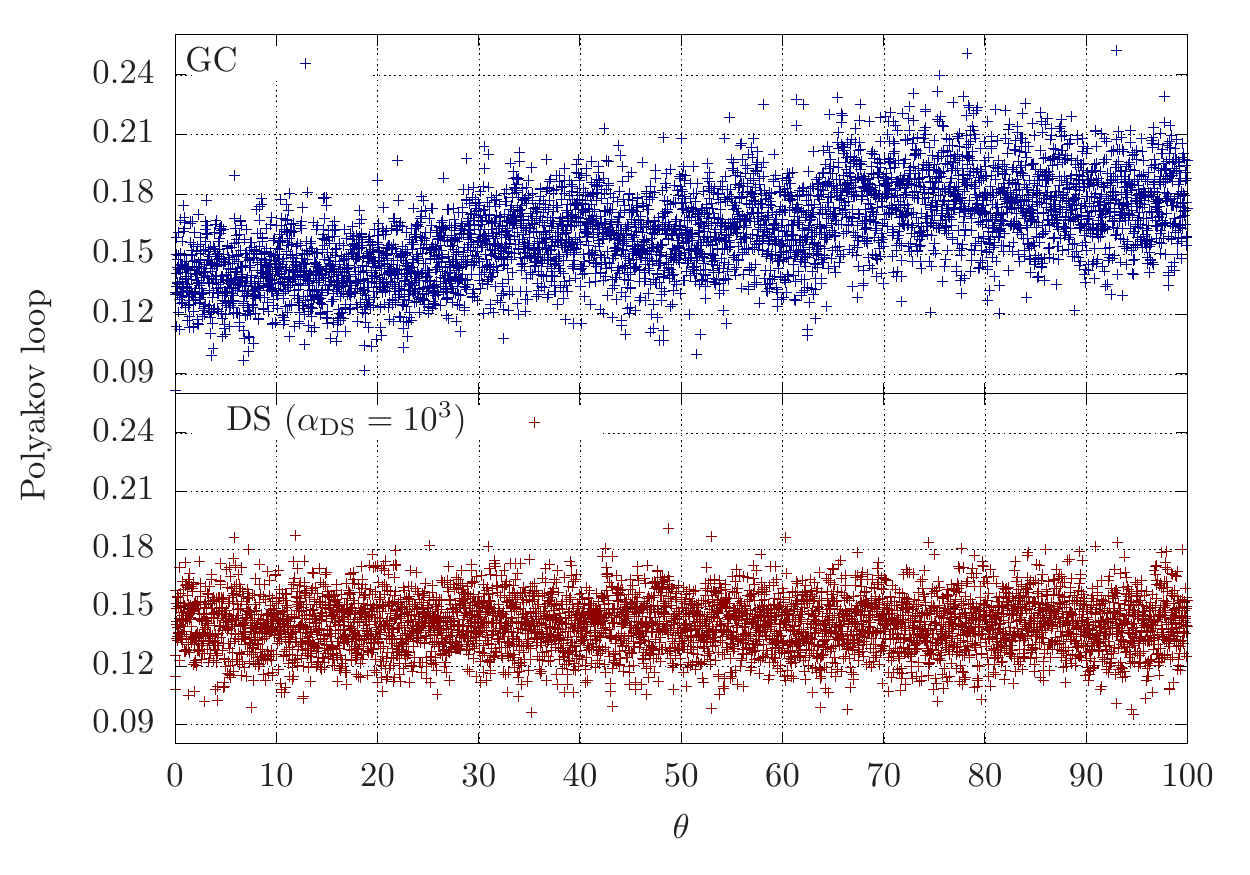}
	\caption{\label{fig.HD.zeroDet.time}The Langevin time evolution of the Polyakov loop
	using gauge cooling (top) and dynamic stabilisation (bottom) for HDQCD on a
	$8^3 \times 20$ lattice, with $\kappa=0.04$, $\beta=5.8$ and $\mu=2.45$.
	After a short Langevin time ($\theta \approx 20$) the Polyakov loop
	changes its behaviour when DS is not used.}
\end{figure}
Figure~\ref{fig.HD.zeroDet.time} shows the Polyakov loop as function of the Langevin time for a scenario with severe sign problem, without and with DS.
These simulations were carried out a volume of $\Omega=8^3 \times 20$ and $\mu=2.45$.
In this case the average phase is $\langle e^{2i\phi} \rangle = -0.0042(35) - 0.0047(35)i$, indicating a very small overlap between the full and phase quenched models.
The Polyakov loop in the simulation
without DS changes to a different value for $\theta \gtrsim 20$ when the unitarity norm
exceeds $O(0.1)$. We use the region before the unitarity norm rises 
as a reference point to test dynamic stabilisation. Due to this small sampling region, the statistical uncertainties of the gauge cooling simulations are comparatively large. The results are compatible for a wide
region of $\alpha_{\mathrm{DS}}$, as shown in fig.~\ref{fig.alpha.scan.det.zero}.
We find disagreement when $\alpha_{\mathrm{DS}}$ is outside a certain window.
This can be understood as follows: for $\alpha_{\mathrm{DS}}$ very small the DS drift is too small to be effective; on the other hand, large values of the control parameter cause a heavy suppression of the exploration of the non-unitary directions.

Figure~\ref{fig.alpha.scan.det.zero} shows the existence of a region in
$\alpha_{\mathrm{DS}}$, which agrees with GC. Our data suggests that
\begin{equation}
	\frac{\partial \mathcal{O}}{\partial \alpha_{\mathrm{DS}}} = 0\,, \label{eq.alphaDS}
\end{equation}
for a given observable $\mathcal{O}$, is a criterion for determining the
region where DS gives the correct values\footnote{We thank Gert Aarts for
suggesting this.}. In other words, the region of least sensitivity to
$\alpha_{\mathrm{DS}}$ provides the best estimate.

\begin{figure}
	\centering
	\includegraphics[width=0.85\linewidth]{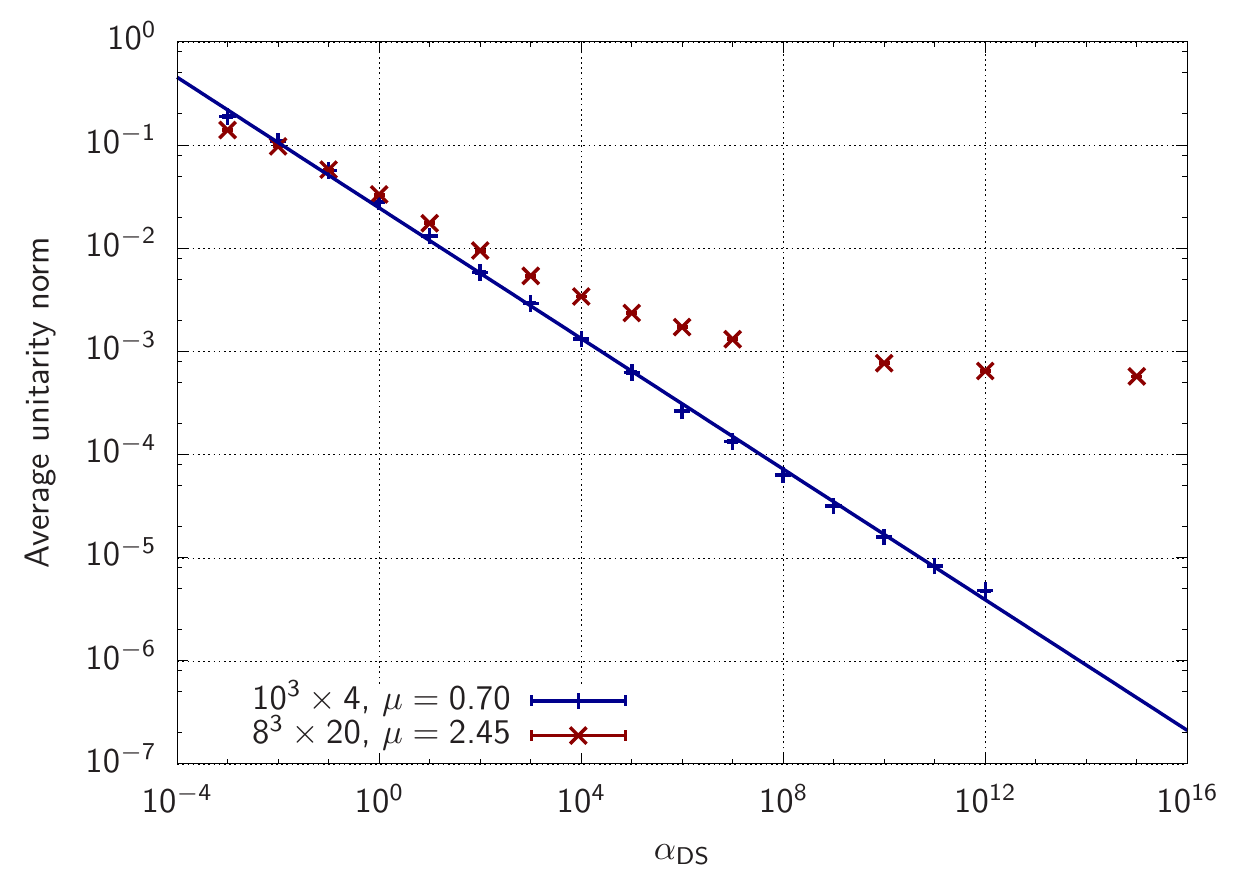}
	\caption{\label{fig.alpha.scan.norm}The average unitarity norm as a function of $\alpha_{\mathrm{DS}}$ for HDQCD, with $\kappa=0.04$, $\beta=5.8$, and volume and chemical potential indicated on the figure.
	A power-law line was added to the data with average phase close to unity to guide the eye.}
\end{figure}

Fig.\ \ref{fig.alpha.scan.norm} shows the average unitarity norm as a function of
$\alpha_{\mathrm{DS}}$ for both previously studied scenarions, with average phases close to unity
and close to zero (parameters shown in the figure).
It is visible that DS is able to restrict the exploration of SL($3,\mathbb{C}$) to submanifolds whose distance to SU($3$) decrease with $\alpha_{\mathrm{DS}}$.
When the sign problem is severe, a plateau in the unitarity norm seems to emerge for large $\alpha_{\mathrm{DS}}$; while for milder sign problems, the results follow a power-law indicated by the line in the figure.

\begin{figure}
	\centering
	\includegraphics[width=0.85\linewidth]{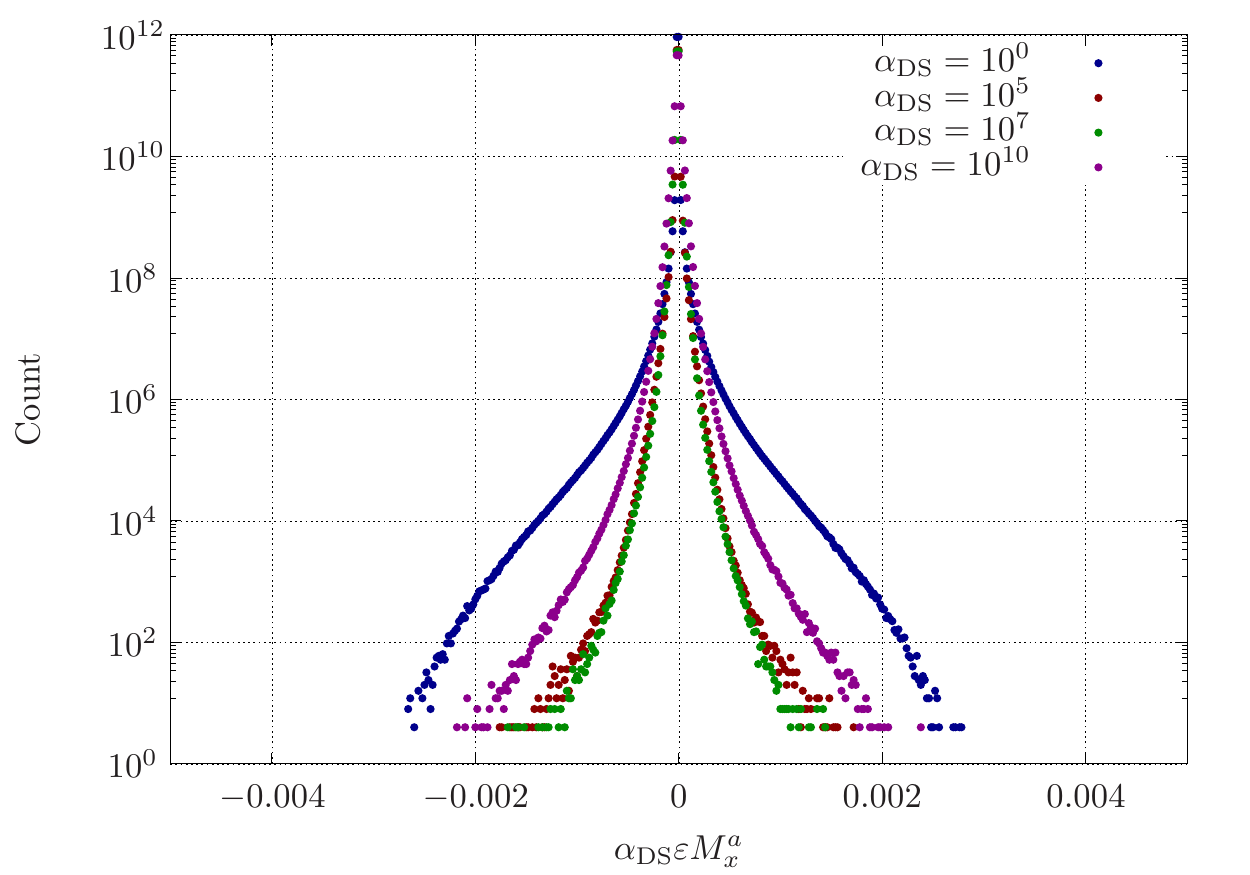}
	\caption{\label{fig.histogram.DS.det.zero}Histogram of the drift added by
	dynamic stabilisation for different values of the control parameter
	$\alpha_{\mathrm{DS}}$.}
\end{figure}

For the remainder of this section we study histograms of the drift of eq.~(\ref{eq.drift.lang.DS}), as they are relevant in the context of the
criteria for correctness~\cite{Aarts:2009uq,Aarts:2011ax,Aarts:2012ft,Aarts:2013uza}: a
heavy-tailed distribution leads to incorrect results.
In fig.~\ref{fig.histogram.DS.det.zero} we show the histograms of the DS drift for four choices for
the control parameter $\alpha_{\mathrm{DS}}$. These simulations correspond to the scenario
with a severe sign problem, i.e. $\Omega=8^3 \times 20$ and $\mu=2.45$. Larger drifts are less frequent than
smaller ones. For large enough values of $\alpha_{\mathrm{DS}}$ the histograms become more localised
distributions.

For $\alpha_{\mathrm{DS}}=10^{0}$ we observe larger values of the product $\alpha_{\mathrm{DS}} \eps M^a_x$, due to the larger unitarity
norm. We remind the reader that $M^a_x$ is a function of the combination $UU^\dagger$ (see eq.
\ref{eq.DS.term}), similar to the unitarity norm.
As $\alpha_{\mathrm{DS}}$ increases, the unitarity norm decreases and then plateaus (seen in
fig.~\ref{fig.alpha.scan.norm}), and so does $M^a_x$.
Intuitively, for very large $\alpha_{\mathrm{DS}}$ the DS drift overshadows the
Langevin drift coming from the physical action.

\begin{figure}
	\centering
	\includegraphics[width=0.85\linewidth]{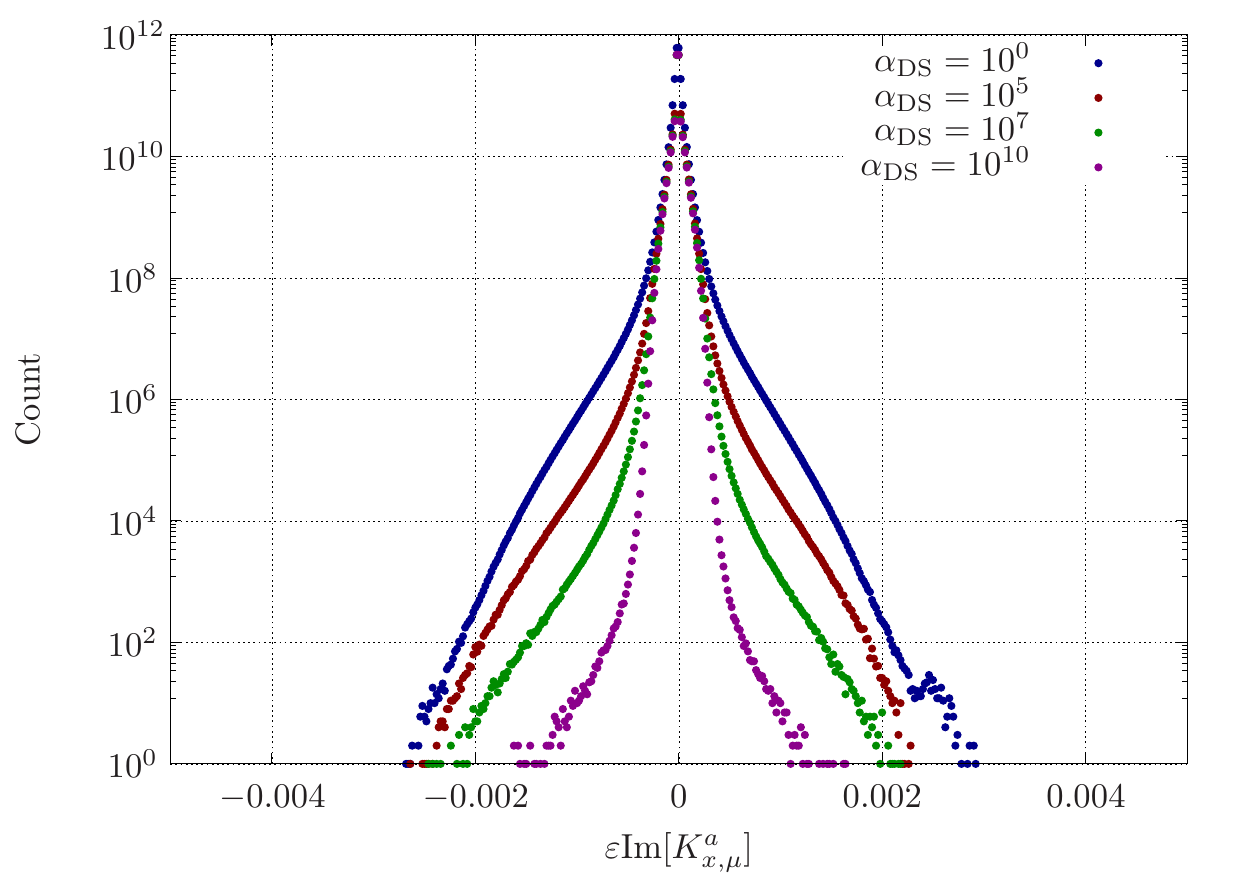}
	\caption{\label{fig.histogram.Real.det.zero}Histogram of the imaginary
	part of the Langevin drift, i.e. eq.~(\ref{eq.im.drift}), multiplied by the
	Langevin step size, for different values of the DS control parameter
	$\alpha_{\mathrm{DS}}$.}
\end{figure}

\begin{figure}
	\centering
	\includegraphics[width=0.85\linewidth]{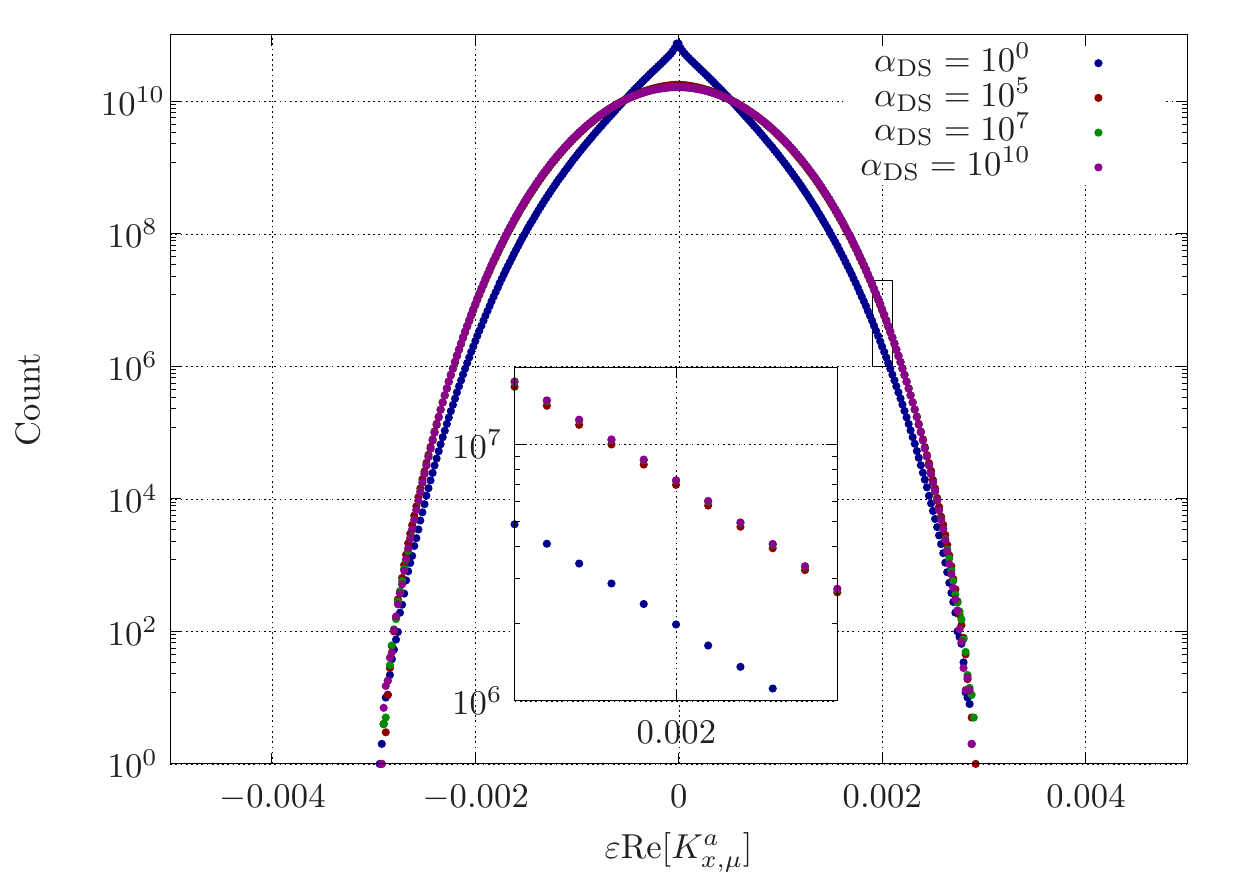}
	\caption{\label{fig.histogram.Imag.det.zero}Histogram of the real
	part of the Langevin drift, i.e. eq.~(\ref{eq.re.drift}), multiplied
	by the Langevin step size, for different values of the DS control parameter $\alpha_{\mathrm{DS}}$.}
\end{figure}

The Langevin drift from eq.~(\ref{eq.drift.lang.DS}) has real and imaginary components given by
\begin{align}
	\text{Re}[K^a_{x,\mu}] &= \text{Re}[-D^a_{x,\mu} S]\,,\label{eq.re.drift} \\ 
	\text{Im}[K^a_{x,\mu}] &= \text{Im}[-D^a_{x,\mu} S] + i \alpha_{\mathrm{DS}} M^a_x\,.\label{eq.im.drift}
\end{align}
The suppression 
of the imaginary part of the Langevin drift can be seen in the histogram of fig.\
\ref{fig.histogram.Real.det.zero}. The real part of the drift is 
plotted in fig.\ \ref{fig.histogram.Imag.det.zero} and essentially remains unchanged by
dynamic stabilisation, once $\alpha_{\mathrm{DS}}$ is sufficiently large, as evident in the inset.
For too small values of $\alpha_{\mathrm{DS}}$, the system can explore a large
region of SL($3,\mathbb{C}$), which causes a different behaviour and lead to convergence to a wrong
limit.

%% file: 7_dsBeta.tex
\section{Continuum behaviour of dynamic stabilisation}\label{sec.betaScan}
\begin{figure}
	\centering
	\includegraphics[width=0.85\linewidth]{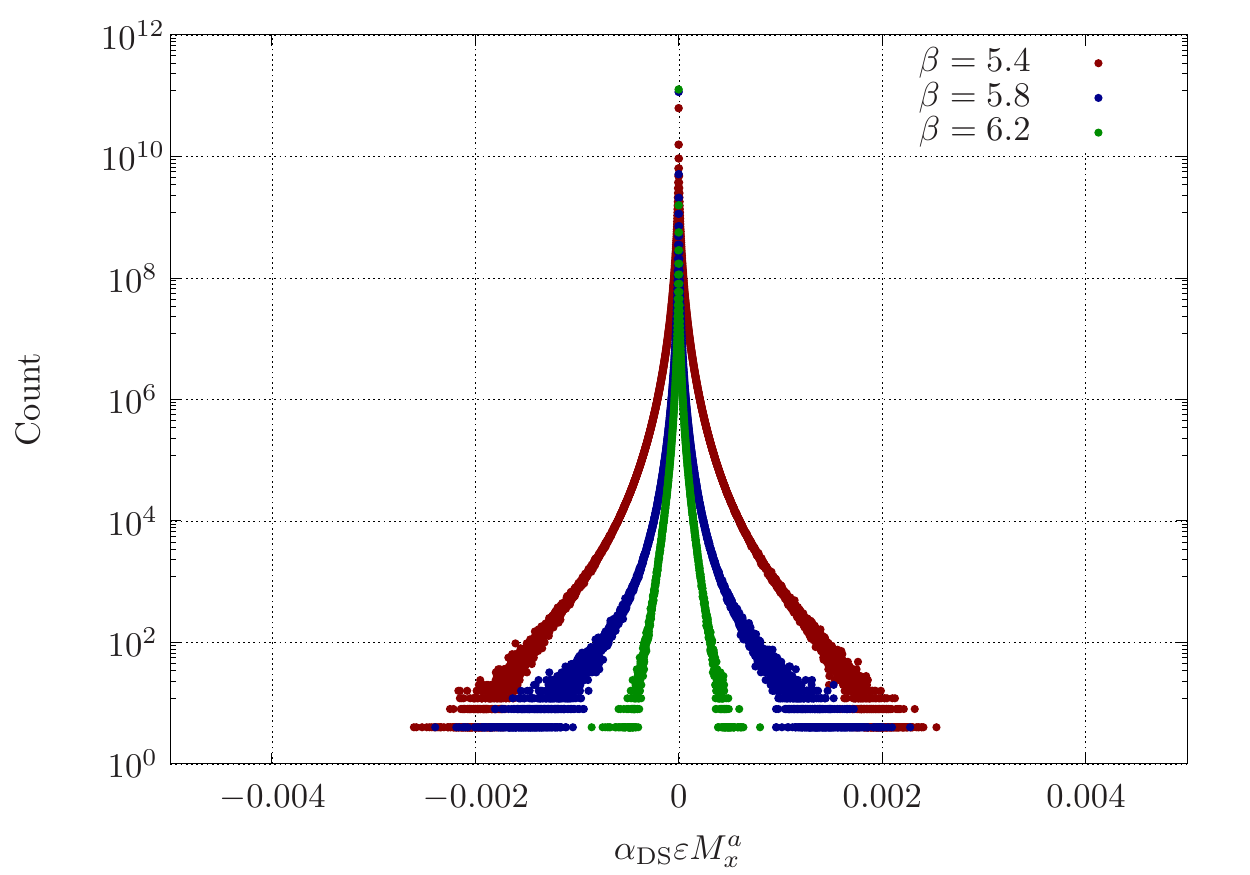}
	\caption{\label{fig.histogram.DS.beta.det.zero}Histogram of the drift added by
	dynamic stabilisation for different values of the gauge coupling $\beta$.}
\end{figure}

\begin{figure}
	\centering
	\includegraphics[width=0.85\linewidth]{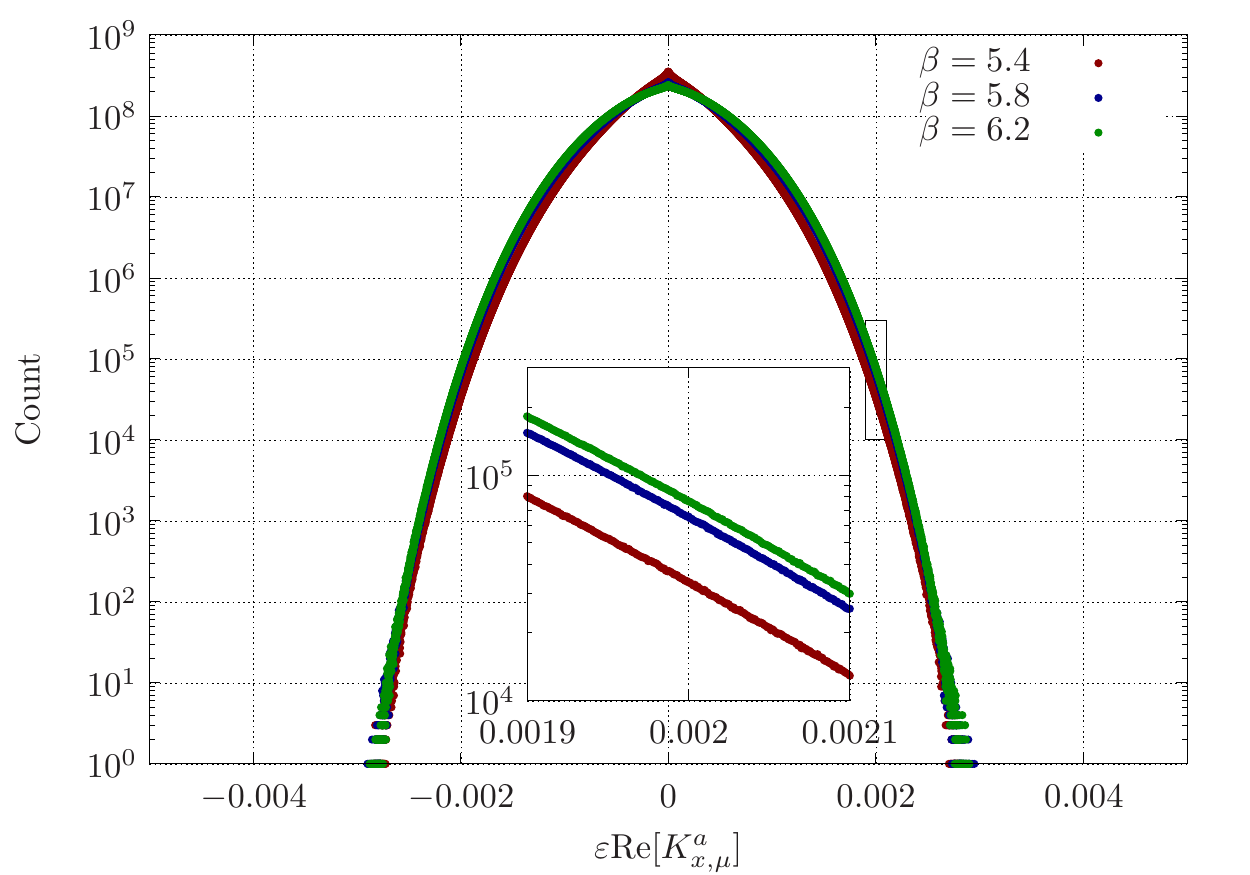}
	\caption{\label{fig.histogram.beta.det.zero}Histogram of the real part of the 
	Langevin drift for different values of the gauge coupling $\beta$.}
\end{figure}

To check the continuum limit of the dynamic stabilisation, we have performed three simulations at
different gauge couplings, specifically $\beta=5.4$, $5.8$ and $6.2$, in a
lattice of volume $8^3 \times 20$, $\kappa=0.04$, $\mu=2.45$ and $\alpha_{\mathrm{DS}} = 10^3$.
As in the previous section, we have applied one gauge cooling step between subsequent Langevin
updates.
The resulting histograms for the DS drift are shown in fig.~\ref{fig.histogram.DS.beta.det.zero}.
At finer lattices, larger values of the dynamic stabilisation drift $M^a_x$ occur less
frequently.
Furthermore, the change in gauge coupling has a small effect on the real part
of the Langevin drift, as seen in fig. \ref{fig.histogram.beta.det.zero}.
These changes correspond to different physics being simulated at different lattice spacings.
This implies that the drift arrising from DS, $M^a_x$, decreases faster than $K^a_x$.

%% file: 8_deconfHD.tex
The deconfinement transition of the heavy dense approximation of QCD was studied in
refs.~\cite{Seiler:2012wz,Aarts:2013uxa}.
The gauge coupling was varied in the interval $5.4 \leq \beta \leq 6.2$. These 
simulations have a lattice of volume $6^3 \times 6$, chemical potential of $\mu=0.85$ and hopping
parameter of $\kappa=0.12$. It was found that the average plaquette from complex Langevin with just gauge cooling disagrees
with reweighting for $\beta \lesssim 5.5$. We have investigated whether dynamic stabilisation
can remedy this discrepancy. As in our previous studies, we added one step of gauge cooling between consecutive
Langevin updates.

\begin{figure}
	\centering
	\includegraphics[width=0.85\linewidth]{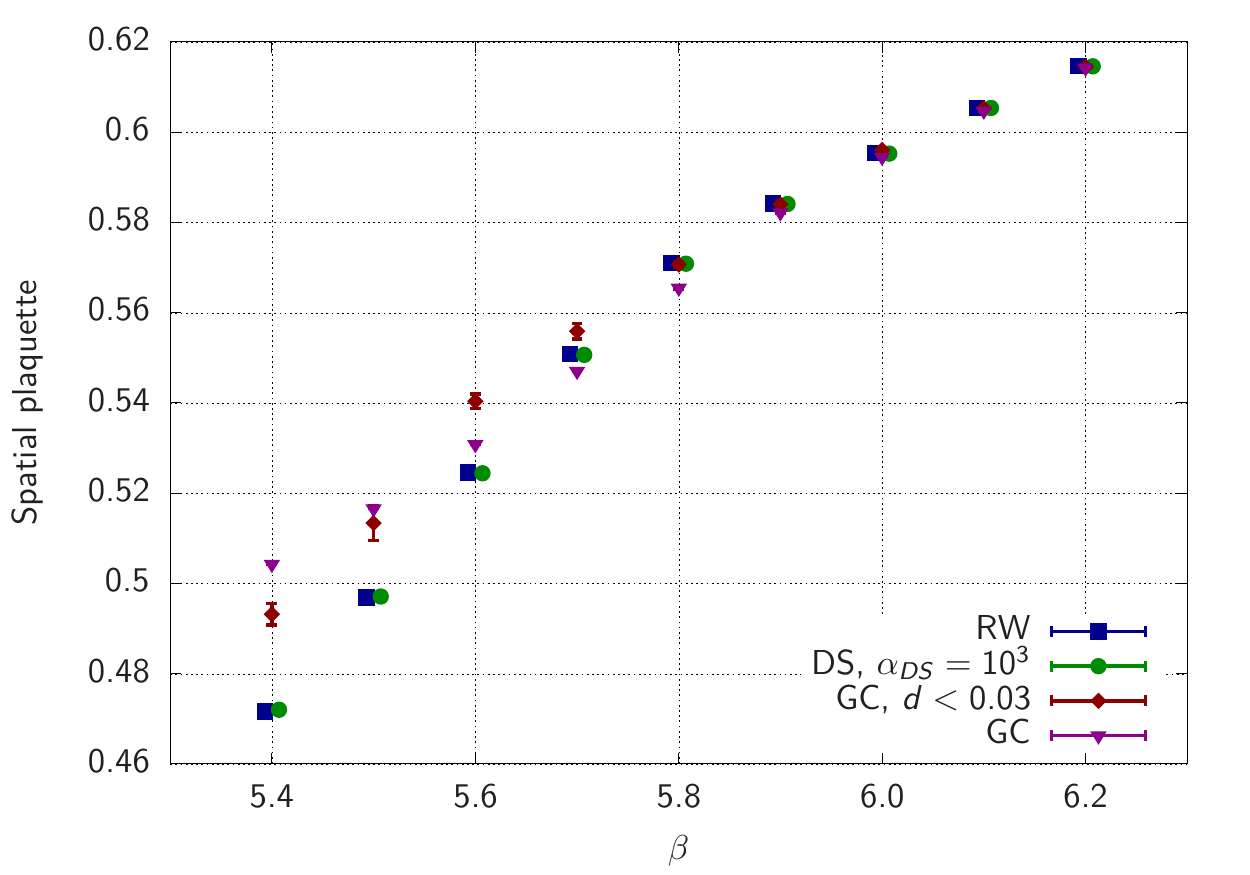} 
	\caption{\label{fig.betaScan.poly}The average spatial plaquette as function of $\beta$.
	Our simulations use a volume $6^3 \times 6$, $\mu=0.85$ and $\kappa=0.12$.
	The data points have been slightly shifted for better readability.}
\end{figure}

Our simulations use an $\alpha_{\mathrm{DS}}=10^3$. 
Figure~\ref{fig.betaScan.poly} shows the spatial plaquette as
function of the gauge coupling. We find good agreement between complex Langevin simulations
using dynamic stabilisation and the reweighting results from
refs.~\cite{Seiler:2012wz,Aarts:2013uxa, NucuPrivate1} in both confined and deconfined phases.
In table~\ref{tb.plaquette.betaScan} we further list the average spatial plaquette between simulations
that used reweighting, dynamic stabilisation and just gauge cooling. The discrepancy between
reweighting and gauge cooling results is clearly visible.
On the other hand, agreement between DS and reweighting can be seen for all
values of $\beta$ used in the simulations.
\begin{table*}
	\centering
\caption{\label{tb.plaquette.betaScan}The average value for the spatial plaquette, from HDQCD
simulations at $6^3 \times 6$, $\kappa=0.12$ and $\mu=0.85$, using reweighting, dynamic
stabilisation and gauge cooling. Reweighting data have been taken from
\protect\cite{Seiler:2012wz,Aarts:2013uxa, NucuPrivate1}}.
\begin{tabular}{cccc}
	\hline
	\noalign{\smallskip}
	$\beta$ & RW & DS & GC \\
	\noalign{\smallskip}
	\hline
	\noalign{\smallskip}
$5.4$ & $0.47164(33)$ & $0.472007(86)$ & $0.504292(75)$\\
$5.5$ & $0.49687(38)$ & $0.49708(11)$ & $0.516607(56)$\\
$5.6$ & $0.52461(47)$ & $0.52441(12)$ & $0.530817(72)$\\
$5.7$ & $0.55086(63)$ & $0.55064(19)$ & $0.547050(97)$\\
$5.8$ & $0.57097(58)$ & $0.570849(69)$ & $0.56547(22)$\\
$5.9$ & $0.58417(47)$ & $0.584086(37)$ & $0.58220(16)$\\
$6.0$ & $0.59533(42)$ & $0.595220(28)$ & $0.594490(67)$\\
$6.1$ & $0.60533(38)$ & $0.605332(24)$ & $0.604713(50)$\\
$6.2$ & $0.61460(36)$ & $0.614567(22)$ & $0.614275(33)$\\
\noalign{\smallskip}
\hline
\end{tabular}
\end{table*}

%% file: 9_staggered.tex
\section{Staggered quarks}\label{sec.stag.DS}
\subsection{Staggered quarks at \texorpdfstring{$\mu = 0$}~}
In order to evaluate the fermionic contribution to the Langevin drift of eq.~(\ref{eq.drift}) we employ a bilinear noise 
scheme and the conjugate gradient method to calculate the trace and inverse, respectively.
One characteristic of the bilinear noise scheme is that, at $\mu=0$, the drift is real only on average~\cite{Sexty:2013ica}.
Therefore, a non-zero unitarity norm is expected even for vanishing chemical potential.
This can cause simulations to diverge. We have investigated whether DS is able to
successfully keep the unitarity norm under control, by comparing complex Langevin and hybrid Monte-Carlo (HMC) simulations\footnote{We thank Philippe de Forcrand for providing the results from hybrid Monte-Carlo simulations.}.
We have used four different lattice volumes, $6^4$, $8^4$, $10^4$ and $12^4$.

First, we have identified a suitable value for the control parameter $\alpha_{\mathrm{DS}}$ following the procedure 
in section~\ref{sec.alphaScan}, i.e. using equation~(\ref{eq.alphaDS}). After finding the optimal
values for $\alpha_{\mathrm{DS}}$ for each lattice size, we extrapolated the results to zero Langevin step size.
We have performed studies with four degenerate quark flavours of mass $m = 0.025$ and inverse
coupling $\beta = 5.6$.
We have analysed the average values of the plaquette and (unrenormalised) chiral condensate.
For these parameters, the quarks are deconfined, indicated by non-zero values for the chiral
condensate.

\begin{figure}
	\centering
	\includegraphics[width=0.85\linewidth]{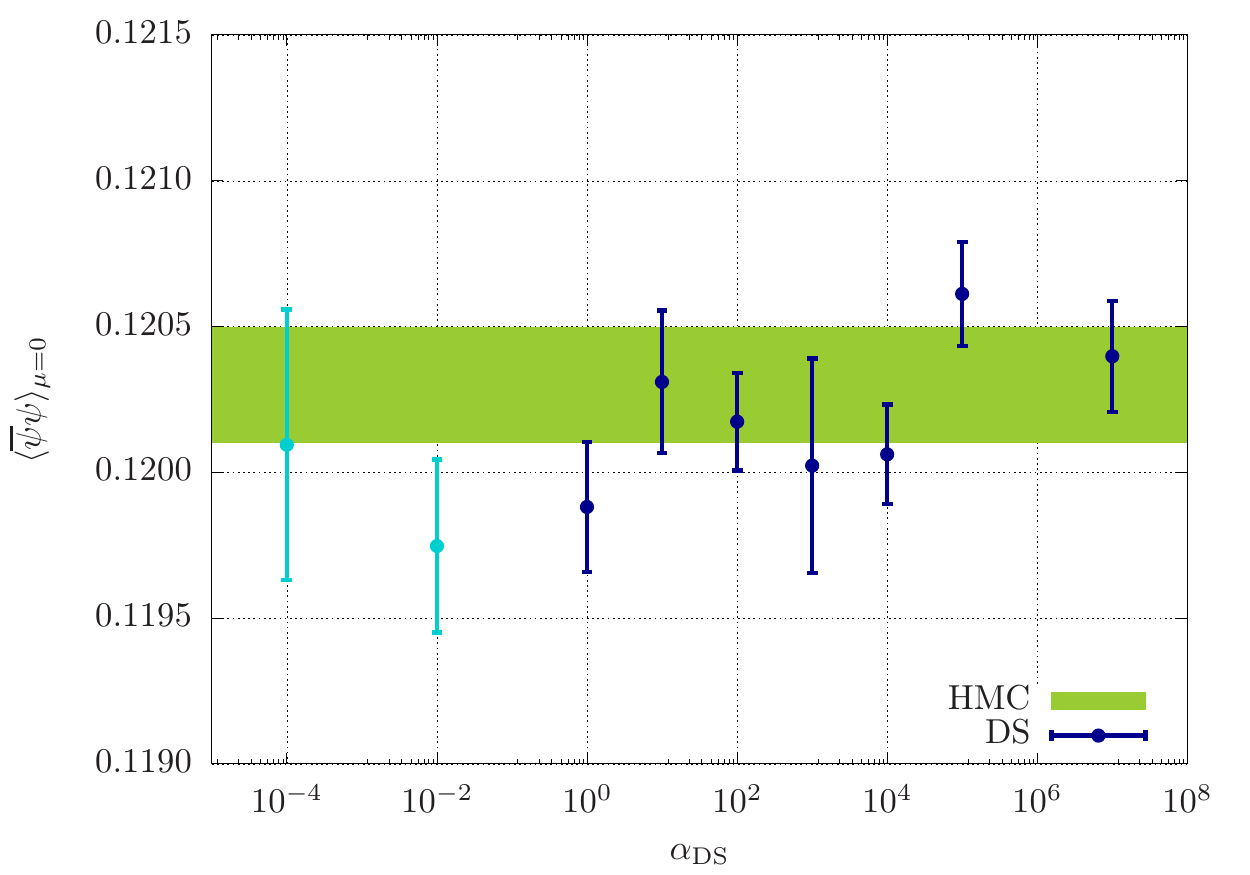} 
	\caption{\label{fig.stag.cc.zeroMu}The chiral condensate at zero chemical potential as a function of
	$\alpha_{\mathrm{DS}}$.
	The green band represents the value obtained from a HMC simulation.
	The simulations were carried out in a volume of $6^3 \times 6$, four quark flavours of mass $m =
	0.025$, and gauge coupling $\beta = 5.6$.}
\end{figure}

The results for the chiral condensate at zero chemical potential for $\Omega=6^4$ are
shown in fig.~\ref{fig.stag.cc.zeroMu}, where the green band 
indicates the result from the HMC run.
The Langevin simulations had an average step size of $\sim 4 \times 10^{-5}$, which leads to approximately 
$3000$--$5000$ independent configurations, including an auto-correlation
analysis as proposed in~\cite{Wolff:2003sm}.
We find very good agreement between hybrid Monte-Carlo and complex Langevin simulations. The two
leftmost points have a unitarity norm larger than $0.03$ and are thus not taken into account.

\begin{figure}
	\centering
	\includegraphics[width=0.85\linewidth]{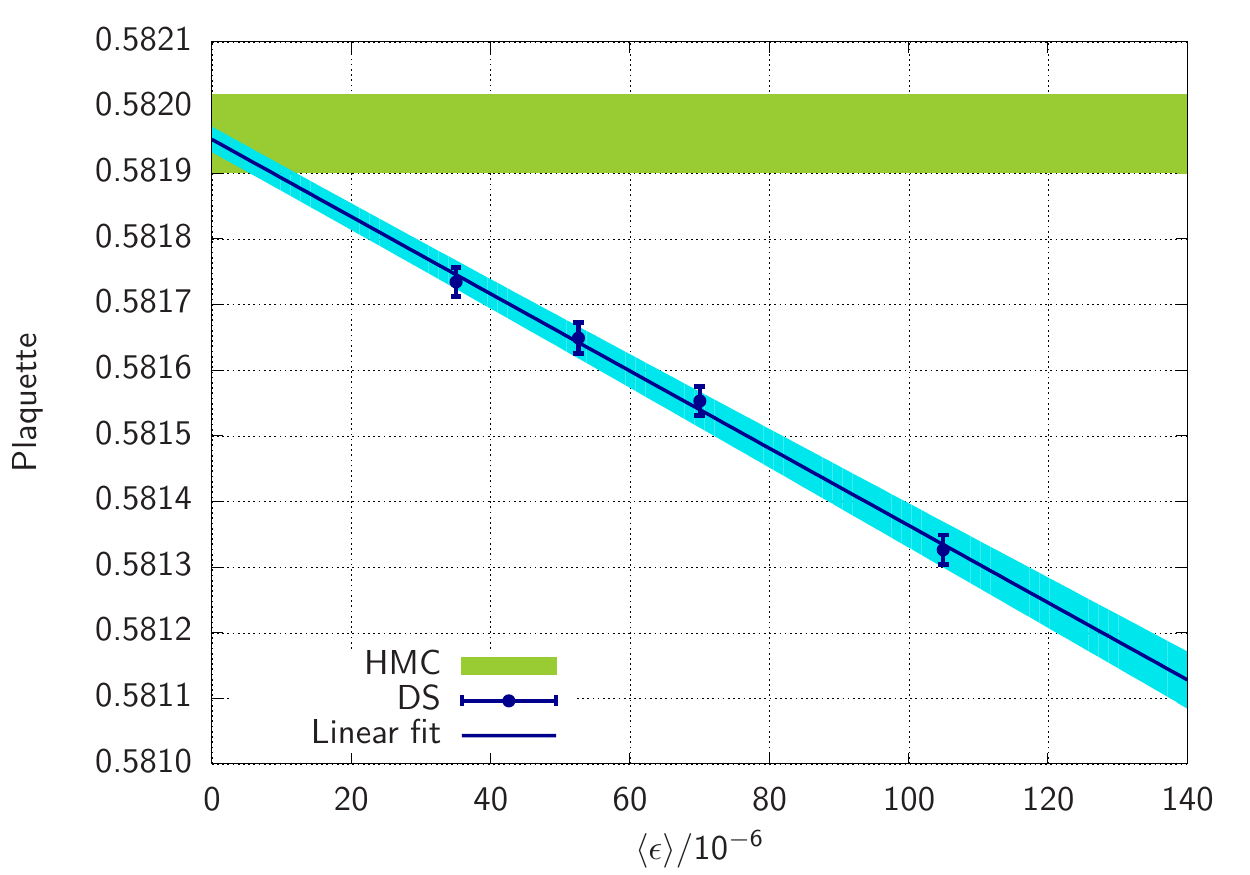}
	\caption{\label{fig.stag.plaq.zeroMu.12.extrap}The plaquette at zero chemical potential as a
	function of the average Langevin step size.
	The green band represents the value obtained from a HMC simulation and the blue region depicts
	the error band from the linear fit to the Langevin data.}
\end{figure}

Figure~\ref{fig.stag.plaq.zeroMu.12.extrap} displays a comparison of the plaquette between
complex Langevin and HMC for the lattice volume of $12^4$.
A straight line has been fitted to the points generated by the Langevin simulations to extrapolate
to zero step size, as the integration scheme is of first order~\cite{Damgaard:1987rr}. 
We find clear agreement within the quoted uncertainties. Results for the average
plaquette and chiral condensate after extrapolation to zero step size can be found in
table~\ref{tb.stag.extrapolated}. The table shows excellent agreement between HMC and CL simulations
for both observables in all four volumes considered.

\begin{table*}
	\centering
\caption{\label{tb.stag.extrapolated}Average values for the plaquette and chiral condensate from
simulations of four flavours of na\"ive staggered fermions at $\beta=5.6$, $m=0.025$ and $\mu=0$. The
Langevin results have been obtained after extrapolation to zero step size.}
\begin{tabular}{ccccc}
	\hline\noalign{\smallskip}
	 & \multicolumn{2}{c}{Plaquette} & \multicolumn{2}{c}{$\overline{\psi} \psi$} \\
	 Volume & HMC & Langevin & HMC & Langevin \\
	\noalign{\smallskip}\hline\noalign{\smallskip}
$6^4$  & $0.58246(8)$ & $0.582452(4)$ & $0.1203(3)$ & $0.1204(2)$\\
$8^4$  & $0.58219(4)$ & $0.582196(1)$ & $0.1316(3)$ & $0.1319(2)$\\
$10^4$ & $0.58200(5)$ & $0.58201(4)$ & $0.1372(3)$ & $0.1370(6)$\\
$12^4$ & $0.58196(6)$ & $0.58195(2)$ & $0.1414(4)$ & $0.1409(3)$\\
\noalign{\smallskip}\hline
\end{tabular}
\end{table*}

Recent works on complex Langevin and gauge cooling applied to staggered 
fermions include~\cite{Sinclair:2015kva, Sinclair:2016nbg}.
There, a discrepancy between the CLE and exact results is reported
for $V=12^4$ at the same inverse coupling and quark mass used here, but with two flavours of staggered fermions.
This tension can potentially be removed by using dynamic stabilization and careful extrapolation to zero step size.
A larger volume ($V=16^4$) has been considered in~\cite{Sinclair:2017zhn}.

\subsection{Staggered quarks at \texorpdfstring{$\mu \neq 0$}~}

We have carried out a qualitative simulation of staggered quarks at high temperatures
spanning a wide range of $\mu$. The chemical potentials vary from $\mu=0$ until saturation, where the entire
lattice is filled with quarks. We use lattices with a spatial volume of $V=12^3$ for two
different temperatures, $N_\tau = 2$ and $4$, with two degenerate quark flavours of mass $m=0.025$.
The inverse coupling is fixed to $\beta=5.6$. With these input parameters, the pion and nucleon
masses are $m_\pi \approx 0.42$ and $m_N \approx 0.93$ in lattice units~\cite{Bitar:1993rk}.
At high temperatures, the inversion of the fermion matrix is numerically cheap and converges quickly
even for large $\mu$. At lower temperatures, however, we have seen that the inversion becomes more expensive, 
as the number of iterations easily exceed $10^4$. Further work using more state-of-the-art
inverters and algorithms are under way and will enable simulations at lower temperatures.
As before, we employ one step of gauge cooling and add a DS force with a control
parameter of $\alpha_{\mathrm{DS}} = 10^3$.
\begin{figure}
\centering
	\includegraphics[width=0.85\linewidth]{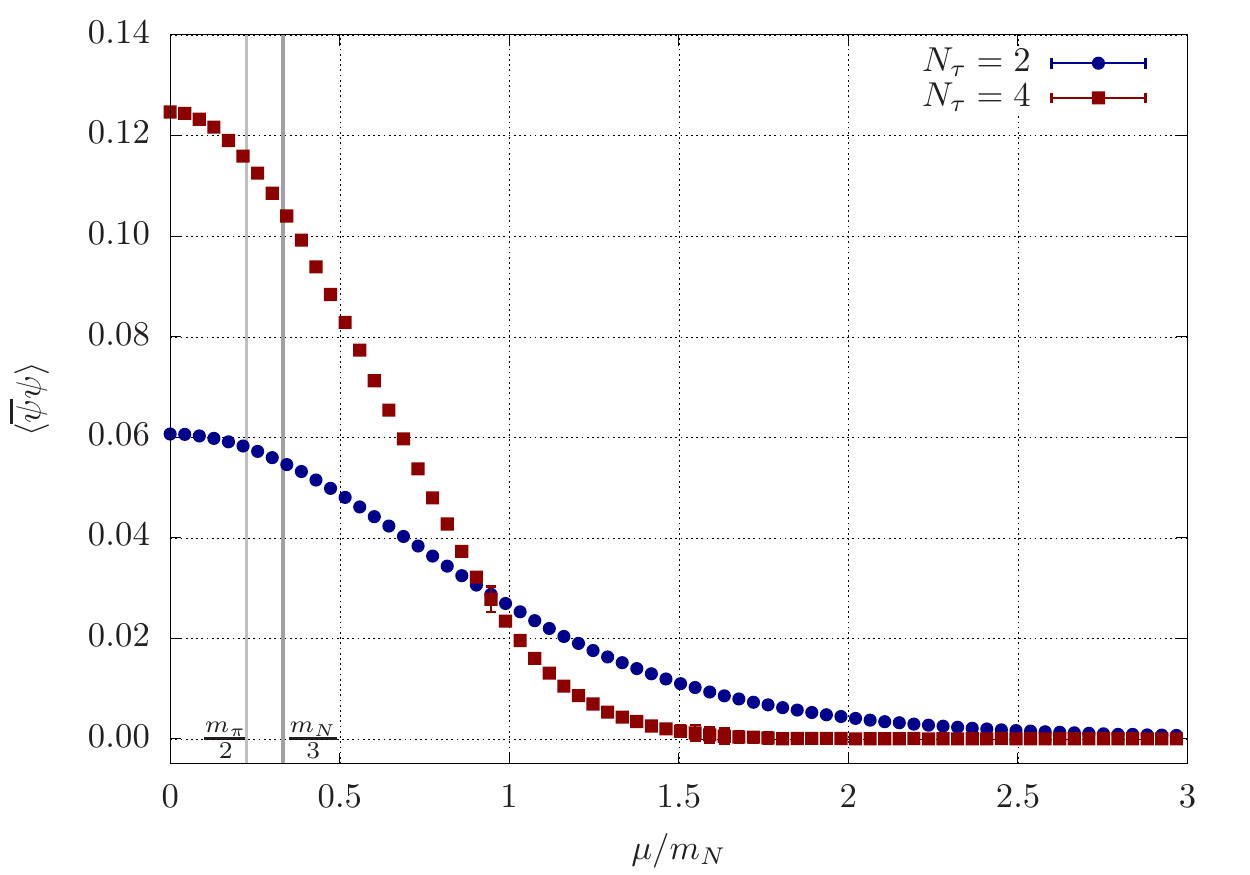}
	\caption{\label{fig.stag.cc1}The chiral condensate as a function of the chemical
	potential, in units of the nucleon mass, for different temperatures.
	Also indicated are the lines of pion (left) and baryon (right) condensation.}
\end{figure}
The chiral condensate, shown in fig.~\ref{fig.stag.cc1},
is not extrapolated to zero step size, but serves a proof of principle.   
Also shown in the plot are vertical lines indicating the regions of pion ($\mu=m_{\pi}/2$) and baryon condensation ($\mu=m_N/3$).
This figure shows that there are, in principle, no obstacles for a complex Langevin simulation of dynamical quarks.

%% file: 10_conclusion.tex
\section{Summary and outlook}\label{sec.summary}
Dynamic stabilisation (DS) was introduced to deal with instabilities found in complex Langevin
simulations, especially when the gauge coupling is small or the unitarity norm rises steadily. The
method is based on adding a non-holomorphic drift to complex Langevin dynamics to keep
simulations in the vicinity of the SU($3$) manifold. We have studied the dependence
of the observables on the control parameter $\alpha_{\mathrm{DS}}$ and have presented a criterion to tune it 
appropriately. We also found numerical evidence that the DS drift decreases when the lattice spacing is
reduced and has a localised distribution.
Dynamic stabilisation improved results on the deconfinement transition for HDQCD, previously
shown in \cite{Seiler:2012wz}, where a discrepancy between reweighting and complex Langevin was observed.
We find good agreement with reweighting for all gauge couplings in both confined and deconfined phases.

We presented a study of complex Langevin simulations of QCD with na\"ive staggered fermions at
vanishing chemical potential. After extrapolating the Langevin results to zero step size, we
found excellent agreement between complex Langevin and hybrid Monte-Carlo
simulations for the plaquette and chiral condensate for four different lattice
volumes, despite dynamic stabilisation adding a non-holomorphic drift. Our findings rectify the
discrepancy found in earlier studies in~\cite{Sinclair:2015kva, Sinclair:2016nbg}. For $\mu > 0$,
we were able to observe changes in the chiral condensate as the chemical potential increases at high temperatures. 
In those cases, dynamic stabilisation kept the unitarity norm under control and allowed for long simulations.
However, the extent of these studies were limited, as
simulations at lower temperatures showed a numerical difficulty arising from the inversion of the
fermion matrix.

More analytical work on the justification of dynamic stabilisation is
desirable, as DS formally violates the proof of convergence of CL.
Nevertheless, numerical
evidence clearly shows no difference between HMC and CL, even at a
sub-permille level. This needs to be confirmed at non-zero $\mu$, by comparing with
other approaches, such as those mentioned in section~\ref{intro}.
We are currently working on using improved algorithms and state-of-the-art inversion techniques.
Those will allow for better control over the inversion of quark matrix and enable
simulations at lower temperatures and finite density.

%% file: 3_hdqcd.tex
\section{Heavy-dense QCD}\label{sec.hdqcd}
A useful testing ground for methods to deal with the sign problem is the heavy-dense limit of QCD
(HDQCD)~\cite{Bender:1992gn,Aarts:2008rr}.
In this model, quarks can only evolve in the (Euclidean) temporal direction, with the spatial
hoppings neglected. The gluonic action is the standard Wilson gauge action.
The temporal hoppings are kept, so that all dependence on the chemical potential is retained.
This model shares interesting features with QCD, such as the sign and overlap problems as well as a phase transition at zero temperature and finite $\mu$, and is
therefore ideal for testing the effects of dynamic stabilisation.
The QCD effective action reads
\begin{equation}
	S = S_{\mathrm{YM}} - \ln \det M(U, \mu)\,,
\end{equation}
with $S_{\mathrm{YM}}$ being the Wilson gauge action, at inverse coupling $\beta$.
The HDQCD fermion determinant simplifies to
\begin{equation}
	\det M = \prod_{\vec{x}} \left\{ \det \left[ 1 + h e^{\mu/T} \mathcal{P}_{\vec{x}} \right]^2 \left[ 1 + h e^{-\mu/T} \mathcal{P}^{-1}_{\vec{x}} \right]^2 \right\}\,,
\end{equation}
where $T = 1/N_\tau$ is the temperature, $\mu$ is the chemical potential, $h = (2
\kappa)^{N_\tau}$. Lattice units are used throughout this paper.
The Polyakov loop and its inverse read
\begin{equation}
	\mathcal{P}_{\vec{x}} = \prod^{N_\tau-1}_{\tau=0} U_{(\vec{x},\tau), \hat{4}}\,, \quad
	\mathcal{P}^{-1}_{\vec{x}} = \prod^{0}_{\tau=N_\tau-1} U^{-1}_{(\vec{x},\tau), \hat{4}}\,.
\end{equation}
The term with $e^{-\mu/T}$, which is irrelevant in the heavy-dense limit ($\kappa \to 0$ and $\mu \to \infty$, with $\kappa e^\mu$ kept constant), is required for the symmetry $\left[ \det M(\mu) \right]^* = \det M(-\mu^*)$.

%% file: 4_staggeredIntro.tex
\section{Staggered quarks}\label{sec.stag}
We have used the unimproved staggered fermion action for $N_F$ fermion flavours, whose matrix
elements read
\begin{align}
	M(U, \mu)_{x,y} &= m \delta_{x,y} + \sum_\nu \frac{\eta_\nu(x)}{2} 
	\left[ e^{\mu \delta_{\nu, 4}} U_{x,\nu} \delta_{x + a_{\nu}, y} \right.\nonumber\\
	&\quad\left. - e^{-\mu \delta_{\nu, 4}} U^{-1}_{x-a_{\nu},\nu} \delta_{x - a_{\nu}, y}
	\right]\,,\label{eq.stag.matrix}
\end{align}
with $x$ and $y$ being spacetime coordinates, $\eta_\nu(x)$ the Smit-Kawamoto phase and $a_\nu$ is
a unit vector in the $\nu$-direction.
\iffalse
,
\begin{equation}
	\eta_\nu(x) = (-1)^{\sum_{i=1}^{\nu-1} x_\mu}\,, \quad \eta_1(x) = 1\,.
\end{equation}
\fi
The staggered fermion matrix has the symmetry,
\begin{equation}
	\epsilon_x M(U,\mu)_{x,y} \epsilon_y = M^*(U, -\mu^*)_{y,x}\,,
\end{equation}
where $\epsilon_x = (-1)^{x_1 + x_2 + x_3 + x_4}$ is the staggered equivalent of $\gamma_5$.
This implies that $\det M(U, \mu) = \det M^\dagger(U, -\mu^*)$, leading to a complex effective
action and a sign problem for real chemical potentials.

The Langevin drift originating from the effective action using eq. (\ref{eq.stag.matrix}) is given by
\begin{align}
	K_F &= D^a_{x,\mu} \ln \det M(U,\mu) \nonumber\\
	&=\frac{N_F}{4} \Tr \left[ M^{-1}(U, \mu) D^a_{x,\mu} M(U, \mu) \right]\,.
\end{align}
We have used the bilinear noise scheme and the conjugate gradient method to evaluate the trace
and inversion, respectively, in the fermionic drift~\cite{Sexty:2013ica}.
At $\mu=0$ the drift is real.
However, in the bilinear noise scheme this is true only on average.
Therefore, a non-zero unitarity norm is expected.

%% file: rev.bbl
\providecommand{\href}[2]{#2}\begingroup\raggedright\begin{thebibliography}{10}

\bibitem{Borsanyi:2013bia}
S.~Borsanyi, Z.~Fodor, C.~Hoelbling, S.~D. Katz, S.~Krieg and K.~K. Szabo,
  ~\href{http://dx.doi.org/10.1016/j.physletb.2014.01.007}{Phys. Lett. {\bf
  B730} (2014) 99--104} [\href{http://arxiv.org/abs/1309.5258}{{1309.5258}}].

\bibitem{Bazavov:2014pvz}
{\scshape HotQCD} collaboration, A.~Bazavov et~al.,
  ~\href{http://dx.doi.org/10.1103/PhysRevD.90.094503}{Phys. Rev. {\bf D90}
  (2014) 094503} [\href{http://arxiv.org/abs/1407.6387}{{1407.6387}}].

\bibitem{deForcrand:2010ys}
P.~de~Forcrand, ~{PoS {\bf LAT2009} (2009) 010}
  [\href{http://arxiv.org/abs/1005.0539}{{1005.0539}}].

\bibitem{Aarts:2015tyj}
G.~Aarts, ~\href{http://dx.doi.org/10.1088/1742-6596/706/2/022004}{J. Phys.
  Conf. Ser. {\bf 706} (2016) 022004}
  [\href{http://arxiv.org/abs/1512.05145}{{1512.05145}}].

\bibitem{Parisi:1984cs}
G.~Parisi, ~\href{http://dx.doi.org/10.1016/0370-2693(83)90525-7}{Phys. Lett.
  {\bf 131B} (1983) 393--395}.

\bibitem{Klauder:1983nn}
J.~R. Klauder, ~\href{http://dx.doi.org/10.1007/978-3-7091-7651-1_8}{Acta Phys.
  Austriaca Suppl. {\bf 25} (1983) 251--281}.

\bibitem{Klauder:1983zm}
J.~R. Klauder, ~\href{http://dx.doi.org/10.1088/0305-4470/16/10/001}{J. Phys.
  {\bf A16} (1983) L317}.

\bibitem{Klauder:1983sp}
J.~R. Klauder, ~\href{http://dx.doi.org/10.1103/PhysRevA.29.2036}{Phys. Rev.
  {\bf A29} (1984) 2036--2047}.

\bibitem{deForcrand:2014tha}
P.~de~Forcrand, J.~Langelage, O.~Philipsen and W.~Unger,
  ~\href{http://dx.doi.org/10.1103/PhysRevLett.113.152002}{Phys. Rev. Lett.
  {\bf 113} (2014) 152002}
  [\href{http://arxiv.org/abs/1406.4397}{{1406.4397}}].

\bibitem{Glesaaen:2015vtp}
J.~Glesaaen, M.~Neuman and O.~Philipsen,
  ~\href{http://dx.doi.org/10.1007/JHEP03(2016)100}{JHEP {\bf 03} (2016) 100}
  [\href{http://arxiv.org/abs/1512.05195}{{1512.05195}}].

\bibitem{deForcrand:2017fky}
P.~de~Forcrand, W.~Unger and H.~Vairinhos,
  ~\href{http://dx.doi.org/10.1103/PhysRevD.97.034512}{Phys. Rev. {\bf D97}
  (2018) 034512} [\href{http://arxiv.org/abs/1710.00611}{{1710.00611}}].

\bibitem{Witten:2010cx}
E.~Witten, ~{AMS/IP Stud. Adv. Math. {\bf 50} (2011) 347--446}
  [\href{http://arxiv.org/abs/1001.2933}{{1001.2933}}].

\bibitem{Cristoforetti:2012su}
{\scshape AuroraScience} collaboration, M.~Cristoforetti, F.~Di~Renzo and
  L.~Scorzato, ~\href{http://dx.doi.org/10.1103/PhysRevD.86.074506}{Phys. Rev.
  {\bf D86} (2012) 074506}
  [\href{http://arxiv.org/abs/1205.3996}{{1205.3996}}].

\bibitem{Alexandru:2015sua}
A.~Alexandru, G.~Basar, P.~F. Bedaque, G.~W. Ridgway and N.~C. Warrington,
  ~\href{http://dx.doi.org/10.1007/JHEP05(2016)053}{JHEP {\bf 05} (2016) 053}
  [\href{http://arxiv.org/abs/1512.08764}{{1512.08764}}].

\bibitem{Fujii:2013sra}
H.~Fujii, D.~Honda, M.~Kato, Y.~Kikukawa, S.~Komatsu and T.~Sano,
  ~\href{http://dx.doi.org/10.1007/JHEP10(2013)147}{JHEP {\bf 10} (2013) 147}
  [\href{http://arxiv.org/abs/1309.4371}{{1309.4371}}].

\bibitem{Nishimura:2017vav}
J.~Nishimura and S.~Shimasaki,
  ~\href{http://dx.doi.org/10.1007/JHEP06(2017)023}{JHEP {\bf 06} (2017) 023}
  [\href{http://arxiv.org/abs/1703.09409}{{1703.09409}}].

\bibitem{Alexandru:2016gsd}
A.~Alexandru, G.~Basar, P.~F. Bedaque, S.~Vartak and N.~C. Warrington,
  ~\href{http://dx.doi.org/10.1103/PhysRevLett.117.081602}{Phys. Rev. Lett.
  {\bf 117} (2016) 081602}
  [\href{http://arxiv.org/abs/1605.08040}{{1605.08040}}].

\bibitem{Langfeld:2012ah}
K.~Langfeld, B.~Lucini and A.~Rago,
  ~\href{http://dx.doi.org/10.1103/PhysRevLett.109.111601}{Phys. Rev. Lett.
  {\bf 109} (2012) 111601}
  [\href{http://arxiv.org/abs/1204.3243}{{1204.3243}}].

\bibitem{Gattringer:2016kco}
C.~Gattringer and K.~Langfeld,
  ~\href{http://dx.doi.org/10.1142/S0217751X16430077}{Int. J. Mod. Phys. {\bf
  A31} (2016) 1643007} [\href{http://arxiv.org/abs/1603.09517}{{1603.09517}}].

\bibitem{Garron:2016noc}
N.~Garron and K.~Langfeld,
  ~\href{http://dx.doi.org/10.1140/epjc/s10052-016-4412-2}{Eur. Phys. J. {\bf
  C76} (2016) 569} [\href{http://arxiv.org/abs/1605.02709}{{1605.02709}}].

\bibitem{Endrodi:2018zda}
G.~Endrodi, Z.~Fodor, S.~D. Katz, D.~Sexty, K.~K. Szabo and C.~Torok, ~
  \href{http://arxiv.org/abs/1807.08326}{{1807.08326}}.

\bibitem{Alexandru:2018fqp}
A.~Alexandru, P.~F. Bedaque, H.~Lamm and S.~Lawrence,
  ~\href{http://dx.doi.org/10.1103/PhysRevD.97.094510}{Phys. Rev. {\bf D97}
  (2018) 094510} [\href{http://arxiv.org/abs/1804.00697}{{1804.00697}}].

\bibitem{Parisi:1980ys}
G.~Parisi and Y.-s. Wu, ~{Sci. Sin. {\bf 24} (1981) 483}.

\bibitem{Aarts:2008wh}
G.~Aarts, ~\href{http://dx.doi.org/10.1103/PhysRevLett.102.131601}{Phys. Rev.
  Lett. {\bf 102} (2009) 131601}
  [\href{http://arxiv.org/abs/0810.2089}{{0810.2089}}].

\bibitem{Aarts:2010gr}
G.~Aarts and K.~Splittorff,
  ~\href{http://dx.doi.org/10.1007/JHEP08(2010)017}{JHEP {\bf 08} (2010) 017}
  [\href{http://arxiv.org/abs/1006.0332}{{1006.0332}}].

\bibitem{Aarts:2011zn}
G.~Aarts and F.~A. James,
  ~\href{http://dx.doi.org/10.1007/JHEP01(2012)118}{JHEP {\bf 01} (2012) 118}
  [\href{http://arxiv.org/abs/1112.4655}{{1112.4655}}].

\bibitem{Ambjorn:1985iw}
J.~Ambjorn and S.~K. Yang,
  ~\href{http://dx.doi.org/10.1016/0370-2693(85)90708-7}{Phys. Lett. {\bf 165B}
  (1985) 140}.

\bibitem{Ambjorn:1986fz}
J.~Ambjorn, M.~Flensburg and C.~Peterson,
  ~\href{http://dx.doi.org/10.1016/0550-3213(86)90605-X}{Nucl. Phys. {\bf B275}
  (1986) 375--397}.

\bibitem{Aarts:2010aq}
G.~Aarts and F.~A. James,
  ~\href{http://dx.doi.org/10.1007/JHEP08(2010)020}{JHEP {\bf 08} (2010) 020}
  [\href{http://arxiv.org/abs/1005.3468}{{1005.3468}}].

\bibitem{Berges:2006xc}
J.~Berges, S.~Borsanyi, D.~Sexty and I.~O. Stamatescu,
  ~\href{http://dx.doi.org/10.1103/PhysRevD.75.045007}{Phys. Rev. {\bf D75}
  (2007) 045007}
  [\href{http://arxiv.org/abs/hep-lat/0609058}{{hep-lat/0609058}}].

\bibitem{Berges:2007nr}
J.~Berges and D.~Sexty,
  ~\href{http://dx.doi.org/10.1016/j.nuclphysb.2008.01.018}{Nucl. Phys. {\bf
  B799} (2008) 306--329} [\href{http://arxiv.org/abs/0708.0779}{{0708.0779}}].

\bibitem{Aarts:2009uq}
G.~Aarts, E.~Seiler and I.-O. Stamatescu,
  ~\href{http://dx.doi.org/10.1103/PhysRevD.81.054508}{Phys. Rev. {\bf D81}
  (2010) 054508} [\href{http://arxiv.org/abs/0912.3360}{{0912.3360}}].

\bibitem{Aarts:2011ax}
G.~Aarts, F.~A. James, E.~Seiler and I.-O. Stamatescu,
  ~\href{http://dx.doi.org/10.1140/epjc/s10052-011-1756-5}{Eur. Phys. J. {\bf
  C71} (2011) 1756} [\href{http://arxiv.org/abs/1101.3270}{{1101.3270}}].

\bibitem{Aarts:2012ft}
G.~Aarts, F.~A. James, J.~M. Pawlowski, E.~Seiler, D.~Sexty and I.-O.
  Stamatescu, ~\href{http://dx.doi.org/10.1007/JHEP03(2013)073}{JHEP {\bf 03}
  (2013) 073} [\href{http://arxiv.org/abs/1212.5231}{{1212.5231}}].

\bibitem{Aarts:2013uza}
G.~Aarts, P.~Giudice and E.~Seiler,
  ~\href{http://dx.doi.org/10.1016/j.aop.2013.06.019}{Annals Phys. {\bf 337}
  (2013) 238--260} [\href{http://arxiv.org/abs/1306.3075}{{1306.3075}}].

\bibitem{Nagata:2016vkn}
K.~Nagata, J.~Nishimura and S.~Shimasaki,
  ~\href{http://dx.doi.org/10.1103/PhysRevD.94.114515}{Phys. Rev. {\bf D94}
  (2016) 114515} [\href{http://arxiv.org/abs/1606.07627}{{1606.07627}}].

\bibitem{Nagata:2018net}
K.~Nagata, J.~Nishimura and S.~Shimasaki,
  ~\href{http://dx.doi.org/10.1007/JHEP05(2018)004}{JHEP {\bf 05} (2018) 004}
  [\href{http://arxiv.org/abs/1802.01876}{{1802.01876}}].

\bibitem{Seiler:2012wz}
E.~Seiler, D.~Sexty and I.-O. Stamatescu,
  ~\href{http://dx.doi.org/10.1016/j.physletb.2013.04.062}{Phys. Lett. {\bf
  B723} (2013) 213--216} [\href{http://arxiv.org/abs/1211.3709}{{1211.3709}}].

\bibitem{Nagata:2015uga}
K.~Nagata, J.~Nishimura and S.~Shimasaki,
  ~\href{http://dx.doi.org/10.1093/ptep/ptv173}{PTEP {\bf 2016} (2016) 013B01}
  [\href{http://arxiv.org/abs/1508.02377}{{1508.02377}}].

\bibitem{Nagata:2016alq}
K.~Nagata, J.~Nishimura and S.~Shimasaki,
  ~\href{http://dx.doi.org/10.1007/JHEP07(2016)073}{JHEP {\bf 07} (2016) 073}
  [\href{http://arxiv.org/abs/1604.07717}{{1604.07717}}].

\bibitem{Bloch:2017sex}
J.~Bloch, J.~Glesaaen, J.~J.~M. Verbaarschot and S.~Zafeiropoulos,
  ~\href{http://dx.doi.org/10.1007/JHEP03(2018)015}{JHEP {\bf 03} (2018) 015}
  [\href{http://arxiv.org/abs/1712.07514}{{1712.07514}}].

\bibitem{Aarts:2014bwa}
G.~Aarts, E.~Seiler, D.~Sexty and I.-O. Stamatescu,
  ~\href{http://dx.doi.org/10.1103/PhysRevD.90.114505}{Phys. Rev. {\bf D90}
  (2014) 114505} [\href{http://arxiv.org/abs/1408.3770}{{1408.3770}}].

\bibitem{Sexty:2013ica}
D.~Sexty, ~\href{http://dx.doi.org/10.1016/j.physletb.2014.01.019}{Phys. Lett.
  {\bf B729} (2014) 108--111}
  [\href{http://arxiv.org/abs/1307.7748}{{1307.7748}}].

\bibitem{Aarts:2016qrv}
G.~Aarts, F.~Attanasio, B.~Jäger and D.~Sexty,
  ~\href{http://dx.doi.org/10.1007/JHEP09(2016)087}{JHEP {\bf 09} (2016) 087}
  [\href{http://arxiv.org/abs/1606.05561}{{1606.05561}}].

\bibitem{Bender:1992gn}
I.~Bender, T.~Hashimoto, F.~Karsch, V.~Linke, A.~Nakamura, M.~Plewnia et~al.,
  ~\href{http://dx.doi.org/10.1016/0920-5632(92)90265-T}{Nucl. Phys. Proc.
  Suppl. {\bf 26} (1992) 323--325}.

\bibitem{Aarts:2008rr}
G.~Aarts and I.-O. Stamatescu,
  ~\href{http://dx.doi.org/10.1088/1126-6708/2008/09/018}{JHEP {\bf 09} (2008)
  018} [\href{http://arxiv.org/abs/0807.1597}{{0807.1597}}].

\bibitem{Aarts:2016qhx}
G.~Aarts, F.~Attanasio, B.~Jäger and D.~Sexty,
  ~\href{http://dx.doi.org/10.5506/APhysPolBSupp.9.621}{Acta Phys. Polon. Supp.
  {\bf 9} (2016) 621} [\href{http://arxiv.org/abs/1607.05642}{{1607.05642}}].

\bibitem{Attanasio:2016mhc}
F.~Attanasio and B.~Jäger, ~{PoS {\bf LATTICE2016} (2016) 053}
  [\href{http://arxiv.org/abs/1610.09298}{{1610.09298}}].

\bibitem{Attanasio:2017rxk}
F.~Attanasio and B.~Jäger,
  ~\href{http://dx.doi.org/10.1051/epjconf/201817507039}{EPJ Web Conf. {\bf
  175} (2018) 07039} [\href{http://arxiv.org/abs/1710.06165}{{1710.06165}}].

\bibitem{Damgaard:1987rr}
P.~H. Damgaard and H.~Huffel,
  ~\href{http://dx.doi.org/10.1016/0370-1573(87)90144-X}{Phys. Rept. {\bf 152}
  (1987) 227}.

\bibitem{Mollgaard:2013qra}
A.~Mollgaard and K.~Splittorff,
  ~\href{http://dx.doi.org/10.1103/PhysRevD.88.116007}{Phys. Rev. {\bf D88}
  (2013) 116007} [\href{http://arxiv.org/abs/1309.4335}{{1309.4335}}].

\bibitem{Nishimura:2015pba}
J.~Nishimura and S.~Shimasaki,
  ~\href{http://dx.doi.org/10.1103/PhysRevD.92.011501}{Phys. Rev. {\bf D92}
  (2015) 011501} [\href{http://arxiv.org/abs/1504.08359}{{1504.08359}}].

\bibitem{Splittorff:2014zca}
K.~Splittorff, ~\href{http://dx.doi.org/10.1103/PhysRevD.91.034507}{Phys. Rev.
  {\bf D91} (2015) 034507}
  [\href{http://arxiv.org/abs/1412.0502}{{1412.0502}}].

\bibitem{Greensite:2014cxa}
J.~Greensite, ~\href{http://dx.doi.org/10.1103/PhysRevD.90.114507}{Phys. Rev.
  {\bf D90} (2014) 114507}
  [\href{http://arxiv.org/abs/1406.4558}{{1406.4558}}].

\bibitem{Aarts:2017vrv}
G.~Aarts, E.~Seiler, D.~Sexty and I.-O. Stamatescu,
  ~\href{http://dx.doi.org/10.1007/JHEP05(2017)044,
  10.1007/JHEP01(2018)128}{JHEP {\bf 05} (2017) 044}
  [\href{http://arxiv.org/abs/1701.02322}{{1701.02322}}].

\bibitem{Aarts:2009dg}
G.~Aarts, F.~A. James, E.~Seiler and I.-O. Stamatescu,
  ~\href{http://dx.doi.org/10.1016/j.physletb.2010.03.012}{Phys. Lett. {\bf
  B687} (2010) 154--159} [\href{http://arxiv.org/abs/0912.0617}{{0912.0617}}].

\bibitem{Aarts:2013uxa}
G.~Aarts, L.~Bongiovanni, E.~Seiler, D.~Sexty and I.-O. Stamatescu,
  ~\href{http://dx.doi.org/10.1140/epja/i2013-13089-4}{Eur. Phys. J. {\bf A49}
  (2013) 89} [\href{http://arxiv.org/abs/1303.6425}{{1303.6425}}].

\bibitem{Aarts:2015hnb}
G.~Aarts, F.~Attanasio, B.~Jäger, E.~Seiler, D.~Sexty and I.-O. Stamatescu,
  ~{PoS {\bf LATTICE2015} (2016) 154}
  [\href{http://arxiv.org/abs/1510.09098}{{1510.09098}}].

\bibitem{Loheac:2017yar}
A.~C. Loheac and J.~E. Drut,
  ~\href{http://dx.doi.org/10.1103/PhysRevD.95.094502}{Phys. Rev. {\bf D95}
  (2017) 094502} [\href{http://arxiv.org/abs/1702.04666}{{1702.04666}}].

\bibitem{Rammelmuller:2017vqn}
L.~Rammelmüller, W.~J. Porter, J.~E. Drut and J.~Braun,
  ~\href{http://dx.doi.org/10.1103/PhysRevD.96.094506}{Phys. Rev. {\bf D96}
  (2017) 094506} [\href{http://arxiv.org/abs/1708.03149}{{1708.03149}}].

\bibitem{NucuPrivate1}
I.-O. Stamatescu. Private communication, 2016.

\bibitem{Wolff:2003sm}
{\scshape ALPHA} collaboration, U.~Wolff,
  ~\href{http://dx.doi.org/10.1016/S0010-4655(03)00467-3,
  10.1016/j.cpc.2006.12.001}{Comput. Phys. Commun. {\bf 156} (2004) 143--153}
  [\href{http://arxiv.org/abs/hep-lat/0306017}{{hep-lat/0306017}}].

\bibitem{Sinclair:2015kva}
D.~K. Sinclair and J.~B. Kogut, ~{PoS {\bf LATTICE2015} (2016) 153}
  [\href{http://arxiv.org/abs/1510.06367}{{1510.06367}}].

\bibitem{Sinclair:2016nbg}
D.~K. Sinclair and J.~B. Kogut, ~{PoS {\bf LATTICE2016} (2016) 026}
  [\href{http://arxiv.org/abs/1611.02312}{{1611.02312}}].

\bibitem{Sinclair:2017zhn}
D.~K. Sinclair and J.~B. Kogut,
  ~\href{http://dx.doi.org/10.1051/epjconf/201817507031}{EPJ Web Conf. {\bf
  175} (2018) 07031} [\href{http://arxiv.org/abs/1710.08465}{{1710.08465}}].

\bibitem{Bitar:1993rk}
K.~M. Bitar et~al., ~\href{http://dx.doi.org/10.1103/PhysRevD.49.6026}{Phys.
  Rev. {\bf D49} (1994) 6026--6038}
  [\href{http://arxiv.org/abs/hep-lat/9311027}{{hep-lat/9311027}}].

\end{thebibliography}\endgroup
